\documentclass[a4paper]{jpconf}
\usepackage{amsmath}
\usepackage{graphicx}
\usepackage{cite}

\newcommand{\KV}{{\mbox{$\kappa \sigma^{2}$}}}
\newcommand{\SD}{{\mbox{$S \sigma$}}}
\newcommand{\VM}{{\mbox{$\sigma^{2}/M$}}}

\newcommand{\sNN}{{{$\sqrt{s_{_{\mathrm{NN}}}}$}}}

\newcommand{ \be }{\begin{equation}}
\newcommand{ \ee }{\end{equation}}

\begin{document}

\title{Unified Description of Efficiency Correction and Error Estimation for Moments of Conserved Quantities in Heavy-Ion Collisions}

\author{Xiaofeng Luo }

\address{Institute of Particle Physics and Key Laboratory of Quark \& Lepton Physics (MOE), Central China Normal University, Wuhan 430079, China }

\ead{xfluo@mail.ccnu.edu.cn}

\begin{abstract}
A unified description of efficiency correction and error estimation is provided for moments of conserved 
quantifies in heavy-ion collisions. Moments and cumulants are expressed in terms of the factorial moments, which can be easily corrected for the efficiency effect.  By deriving the covariance between factorial moments, one can obtain the general error formula for the efficiency corrected moments based on the error propagation derived from the Delta theorem.  The skellam distribution based Monto Carlo simulation is used to test the Delta theorem and Bootstrap error estimation methods. The statistical errors calculated from the two methods can well reflect the statistical fluctuations of the efficiency corrected moments.

\end{abstract}

\section{Introduction}
In the conjectured two dimensional Quantum Chromodynamics (QCD) phase diagram (Temperature ($T$) versus Baryon chemical potential ($\mu_{B}$)), the QCD Critical Point (CP) at finite $T$ and $\mu_{B}$ is the endpoint of the first order phase transition boundary and a second order phase transition point between hadronic  and de-confined Quark-Gluon Plasma (QGP) phase in the QCD phase diagram~\cite{QCP_Prediction}. Experimental confirmation of a CP will be an excellent test of QCD theory in the non-perturbative region and a milestone of exploring the QCD phase structure. This is one of the main goals of the Beam Energy Scan (BES) program at Relativistic Heavy Ion Collider (RHIC)~\cite{bes}. During the last five years, we have extensively studied the higher moments of conserved quantities distributions in heavy-ion collisions, such net-baryon, net-charge and net-strangeness,  experimentally and theoretically to search for the possible QCD critical point  in the phase diagram of strongly interacting matters~\cite{STAR_BES_PRL,netcharge_PRL,qcp_signal,ratioCumulant,Neg_Kurtosis,qcp_Rajiv,QM2014_baseline,baseline_PRC,HRG_Karsch,SQM2009,HRG_fjh,HRG_baseline,HRG_Nahrgang,Asakawa_formula,chiralfluid,CPOD2013,kenji,PBM_netpdis,PBM_netq,BFriman_EPJC,Asakawa_baseline}. Because it is expected that the higher moments of conserved quantities are sensitive to the correlation length ($\xi$) of the hot dense nuclear matter~\cite{qcp_signal,ratioCumulant,Neg_Kurtosis,Asakawa}.  On the other hand,  cumulants of conserved quantities are direct connection to the susceptibilities of the dynamical system~\cite{science,Lattice}. This allows us to probe bulk properties, such as chemical freeze-out, of the created nuclear matter in heavy-ion collisions by comparing experimental data with theoretical calculations, such as Lattice QCD and Hadron Resonance Gas (HRG) model~\cite{freezeout,BFriman_EPJC}.  

Theoretically, it is predicted that the net-proton number fluctuation is a good proxy
of the fluctuation of net-baryon number in measuring critical fluctuations near the QCD critical point~\cite{Hatta}.  However, if one considers the effects of the non-critical contributions associated with the 
heavy-ion collision dynamics, such as hadronic scattering, resonance decay and baryon number conservation, there could have discrepancies between the net-proton and net-baryon fluctuations~\cite{Asakawa_formula,QM2014_baseline}. Thus, careful studies are needed to understand the non-critical contributions on the observable to search for the CP in heavy-ion collisions.

The STAR Collaboration has published the energy dependence of moments of net-proton, net-charge multiplicity distributions in Au+Au collisions
at {\sNN}=7.7, 11.5, 19.6, 27, 39, 62.4 and 200 GeV~\cite{STAR_BES_PRL,netcharge_PRL}. 
Those data are taken from the first phase of the RHIC BES in the year 2010 and 2011.  In the year 2014,  the experimental data of Au+Au collisions at {\sNN}=14.5 GeV are successfully taken, which can fill in the large $\mu_{B}$ gap between 11.5 and 19.6 GeV.  In the moment analysis, several techniques ~\cite{technique,Delta_theory} have been applied, such as centrality bin width correction (CBWC), defining a new collision centrality with charged particle multiplicity within large pseudo-rapidity ($|\eta| < 1$) and efficiency correction. These techniques are used to address the effects of volume fluctuation,  auto-correlation and finite detector efficiency.

In heavy-ion collision experiments, particle detectors have a finite particle detection efficiency (less than 100 percent),  which is simply resulting from the limited capability for the detector to register the incoming particles. This efficiency effect will lead to loss of particles multiplicity in each event, which will change the shape of the original event-by-event particle multiplicity distributions.  Since higher moments are very sensitive to the shape of the distributions, especially the tails, the moments values could be significantly modified by the detector efficiency, which could distort and/or suppress the original signal induced by the CP. To obtain precise moment measurements for the CP search in heavy-ion collisions, it is important to recover the moments of the original multiplicity distributions with the measured ones by applying an efficiency correction technique.  The efficiency correction is not only important for the values of the moments but also for the statistical errors.  As the moment analysis is statistics hungry analysis, it is crucial to correctly calculate the statistical errors with limited statistics.  Two error estimation methods have been applied to evaluate the statistical errors for the moment analysis in heavy-ion collisions, one is the Bootstrap and the other is the so called Delta theorem, both of which have been proved to be correct and discussed in the reference~\cite{technique,Delta_theory}. However, the Bootstrap method is a computer intensive resampling method and its calculation speed is quit slow comparing with the error calculations done by analytical formulae. To improve the calculation speed, one wants to derive a set of analytical formulae to calculate the statistical errors for efficiency corrected moments based on the Delta theorem. More important, it can provide us a unified description of the efficiency correction and error estimation for moment analysis in heavy-ion collisions. This description is universal and can be applied to data analysis for different experiments.

In this paper, I will focus on discussing the unified description of efficiency correction and error estimation for moment analysis in heavy-ion collisions.  The paper is organised  as follows. In the section 2, we will discuss the efficiency correction in the moment analysis.  The Delta theorem in statistics will be discussed in the section 3 and applied to derive the error formulae for the efficiency corrected moments. In the section 4, we use Monte Carlo simulation to demonstrate the validity of the error formula. Finally, the summary is presented in the section 5.

\section{Efficiency Correction for Moment Analysis in Heavy-Ion Collisions }
The detection efficiency of particles in heavy-ion collisions can be obtained from the so called Monte Carlo (MC) embedding techniques~\cite{spectra_STAR} in heavy-ion collisions. First, simulated tracks are blended into real events at the raw data level. The tracks are propagated through the full simulation of the detector and geometry with a realistic simulation of the detector response. The efficiency can be obtained with small uncertainties (relative errors $\sim 5\%$) by the ratio of matched MC tracks to input MC tracks. It contains the net effects of tracking efficiency, detector acceptance, decays, and interaction losses. The efficiency correction can be easily applied for the particle yield measurements in heavy-ion collisions. However, it is not 
straightforward to get the efficiency corrected results for higher moments of particle multiplicity distributions.
There are many discussions about the efficiency correction methods for moment analysis~\cite{voker_eff1,voker_eff2,STAR_BES_PRL}. The basic idea is to treat the response function of the finite detecting efficiency as a binomial probability distribution and various order efficiency corrected moments can be expressed in terms of the factorial moments, which can be easily corrected for efficiency. 
In the following, we focus on discussing the net-proton moment analysis in heavy-ion collisions. Experimentally, we measure net-proton number event-by-event wise,
$n=n_{p}-n_{\bar{p}}$, which is proton number minus anti-proton number.  The average value over the whole event ensemble is denoted by $<n>$, where the single angle brackets are used to indicate ensemble average of an event-by-event distributions. For simplify, let us discuss constant efficiency case for (anti-)proton within the entire phase space. The probability distribution function of 
measured proton number $n_{p}$ and anti-proton number $n_{\bar{p}}$ can be expressed as~\cite{voker_eff1}:

\begin{equation}  \label{eq:conv} 
\begin{split}
 p({n_p},{n_{\bar p}}) &= \sum\limits_{{N_p} = n_p}^\infty  {\sum\limits_{{N_{\bar p}} = n_{\bar p}}^\infty  {P({N_p},{N_{\bar p}}) \times \frac{{{N_p}!}}{{{n_p}!\left( {{N_p} - {n_p}} \right)!}}{{({\varepsilon _p})}^{{n_p}}}{{(1 - {\varepsilon _p})}^{{N_p} - {n_p}}}} } \\
& \times  \frac{{{N_{\bar p}}!}}{{{n_{\bar p}}!\left( {{N_{\bar p}} - {n_{\bar p}}} \right)!}}{({\varepsilon _{\bar p}})^{{n_{\bar p}}}}{(1 - {\varepsilon _{\bar p}})^{{N_{\bar p}} - {n_{\bar p}}}} 
\end{split}
\end{equation}
where the $P({N_p},{N_{\bar p}})$ is the original joint probability distribution of number of proton ($N_p$) and anti-proton ($N_{\bar p}$), $\varepsilon _p$ and $\varepsilon _{\bar p}$ are the efficiency of proton and anti-proton, respectively. To derive the efficiency correction formula for moments and cumulants, let us introduce the bivariate factorial moments:

\begin{align}
& {F_{i,k}(N_p, N_{\bar p})} = \left \langle \frac{{{N_p}!}}{{\left( {{N_p} - i} \right)!}}\frac{{{N_{\bar p}}!}}{{\left( {{N_{\bar p}} - k} \right)!}}\right \rangle = \sum\limits_{{N_p} = i}^\infty  {\sum\limits_{{N_{\bar p}} = k}^\infty  {P({N_p},{N_{\bar p}})\frac{{{N_p}!}}{{\left( {{N_p} - i} \right)!}}\frac{{{N_{\bar p}}!}}{{\left( {{N_{\bar p}} - k} \right)!}}} }   \label{eq:fact1} \\ 
&  {f_{i,k}(n_p, n_{\bar p})} =\left \langle \frac{{{n_p}!}}{{\left( {{n_p} - i} \right)!}}\frac{{{n_{\bar p}}!}}{{\left( {{n_{\bar p}} - k} \right)!}}\right\rangle  = \sum\limits_{{n_p} = i}^\infty  {\sum\limits_{{n_{\bar p}} = k}^\infty  {p({n_p},{n_{\bar p}})\frac{{{n_p}!}}{{\left( {{n_p} - i} \right)!}}\frac{{{n_{\bar p}}!}}{{\left( {{n_{\bar p}} - k} \right)!}}} }  \label{eq:fact2}
\end{align}

With the Eq. (\ref{eq:conv}), (\ref{eq:fact1}) and (\ref{eq:fact2}), one can obtain a useful relation between the efficiency corrected and uncorrected factorial moments as:
\begin{equation}  \label{eq:relation} 
{F_{i,k}(N_p, N_{\bar p})} = \frac{{{f_{i,k}(n_p, n_{\bar p})}}}{{{{({\varepsilon _p})}^i}{{({\varepsilon _{\bar p}})}^k}}}
\end{equation}

The various order moments and cumulants can be expressed in terms of the factorial moments. Before deriving the formula for the moments and cumulants of net-proton distributions, we need some mathematical relationships between moments, central moments, cumulants and factorial moments. Let us define a multivariate random vector $\boldsymbol{X}=(X_1,X_2,...,X_k)^{'}$ and a set of number $\boldsymbol{r}=(r_1,r_2,...,r_k)^{'}$. The multivariate moments, central moments and factorial moments can be written as:
\begin{align} \label{eq:momdefinition}
& {m_{\boldsymbol{r}}}(\boldsymbol{X}) = E\left[\prod\limits_{i = 1}^k {X_i^{{r_i}}} \right] \\
& {\mu _{\boldsymbol{r}}}(\boldsymbol{X}) = E\left[\prod\limits_{i = 1}^k {(X_i  - E[{X_i}])^{{r_i}}}\right] \\
& {F_{\boldsymbol{r}}}(\boldsymbol{X}) = E\left[\prod\limits_{i = 1}^k {\frac{{X_i^{}!}}{{(X_i^{} - {r_i})!}}}\right]
\end{align}
where $E$ denotes the expectation value operator, and the ${m_{\boldsymbol{r}}}(\boldsymbol{X})$, $ {\mu _{\boldsymbol{r}}}(\boldsymbol{X})$ and  $ {F_{\boldsymbol{r}}}(\boldsymbol{X}) $ are multivariate moments, central moments and factorial moments, respectively. Then, we have the relation between the moments and central moments by using binomial expansions: 
\begin{equation} 
\begin{array}{l}
{\mu _{\boldsymbol{r}}}({\boldsymbol{X}}) = \begin{array}{*{20}{c}}
{\sum\limits_{{i_1} = 0}^{{r_1}} {} }& \cdots &{\sum\limits_{{i_k} = 0}^{{r_k}} {{{( - 1)}^{{i_1} + {i_2} \cdots  + {i_k}}}(\begin{array}{*{20}{c}}
{{r_1}}\\
{{i_1}}
\end{array}) \cdots (\begin{array}{*{20}{c}}
{{r_k}}\\
{{i_k}}
\end{array})} }
\end{array}\\
\begin{array}{*{20}{c}}
{}&{ \times {{(E[{X_1}])}^{{i_1}}} \cdots }&{{{(E[{X_k}])}^{{i_k}}}}
\end{array}m_{{\boldsymbol{r-i}}}^{}({\boldsymbol{X}})
\end{array}
\end{equation} 
where  $\boldsymbol{i}=(i_1,i_2,...,i_k)^{'}$. To get the relation between moments and factorial moments, one needs the Stirling numbers of the first ($s_{1}(n,i)$) and second kind ($s_{2}(n,i)$), which are defined as: 
 \begin{align}
&\frac{{N!}}{{(N - n)!}} = \sum\limits_{i = 0}^n {{s_1}(n,i)} {N^i}\\
&{N^n} = \sum\limits_{i = 0}^n {{s_2}(n,i)} \frac{{N!}}{{(N - i)!}}
\end{align}
 where $N$, $n$ and $i$ are non-negative integer number.
The recursion equations for the Stirling numbers of the first and second kind are:
\begin{equation}
\begin{split}
&{s_1}(n,i) = {s_1}(n - 1,i - 1) - (n - 1)\times{s_1}(n - 1,i)\\
&{\left. {{s_1}(n,i)} \right|_{n < i}} = 0,{\left. {{s_1}(n,i)} \right|_{n = i}} = 1,{\left. {{s_1}(n,0)} \right|_{n > 0}} = 0
\end{split}
\end{equation}
and
\begin{equation}
\begin{split}
&{s_2}(n,i) = {s_2}(n - 1,i - 1) + i\times{s_2}(n - 1,i)\\
&{\left. {{s_2}(n,i)} \right|_{n < i}} = 0,{\left. {{s_2}(n,i)} \right|_{n = i}} = 1,{\left. {{s_2}(n,0)} \right|_{n >0}} = 0 
\end{split}
\end{equation}
The Stirling number of the first kind may have the negative value while the value of the second kind is always non-negative.
With the two kinds of Stirling numbers, one can write down the relations between moments and factorial moments as:
\begin{equation} \label{eq:mtof}
{m_{\boldsymbol{r}}}({\boldsymbol{X}}) = \begin{array}{*{20}{c}}
{\sum\limits_{{i_1} = 0}^{{r_1}} {} }& \cdots &{\sum\limits_{{i_k} = 0}^{{r_k}} {{s_2}({r_1},{i_1})} }
\end{array} \cdots {s_2}({r_k},{i_k}){F_{\boldsymbol{r}}}({\boldsymbol{X}})
\end{equation}

\begin{equation} \label{eq:ftom}
{F_{\boldsymbol{r}}}({\boldsymbol{X}}) = \begin{array}{*{20}{c}}
{\sum\limits_{{i_1} = 0}^{{r_1}} {} }& \cdots &{\sum\limits_{{i_k} = 0}^{{r_k}} {{s_1}({r_1},{i_1})} }
\end{array} \cdots {s_1}({r_k},{i_k}){m_{\boldsymbol{r}}}({\boldsymbol{X}})
\end{equation}

With Eq. (\ref{eq:momdefinition}) to (\ref{eq:ftom}), one can express the moments
of net-proton distributions in terms of the factorial moments. There are two variables in net-proton number calculation, the number of protons ($N_p$) and anti-protons ($N_{\bar{p}}$).
The $n^{th}$ order moments of net-proton distributions can be expressed in term of factorial moments: 

\begin{equation} \label{eq:mtof2}
\begin{array}{l}
{m_n}({N_p} - {N_{\bar p}}) = <{({N_p} - {N_{\bar p}})^n}> = \sum\limits_{i = 0}^n {{{( - 1)}^i}\left( {\begin{array}{*{20}{c}}
n\\
i
\end{array}} \right)} <N_p^{n - i}N_{\bar p}^i>\\
 = \sum\limits_{i = 0}^n {{{( - 1)}^i}\left( {\begin{array}{*{20}{c}}
n\\
i
\end{array}} \right)} \left[ {\sum\limits_{{r_1} = 0}^{n - i} {\sum\limits_{{r_2} = 0}^i {{s_2}(n - i,{r_1}){s_2}(i,{r_2}){F_{{r_1},{r_2}}}({N_p},{N_{\bar p}})} } } \right]\\
 = \sum\limits_{i = 0}^n {\sum\limits_{{r_1} = 0}^{n - i} {\sum\limits_{{r_2} = 0}^i {{{( - 1)}^i}\left( {\begin{array}{*{20}{c}}
n\\
i
\end{array}} \right){s_2}(n - i,{r_1}){s_2}(i,{r_2}){F_{{r_1},{r_2}}}({N_p},{N_{\bar p}})} } } 
\end{array}
\end{equation}
Actually, two steps are needed to obtain this equation, the first step is to expand the moments of net-proton to the bivariate moments by using binomial expansion, and the other one is 
to express the bivariate moments in term of the factorial moments using the Eq. (\ref{eq:mtof}). Now, one can easily calculate the efficiency corrected moments of net-proton distributions in heavy-ion collisions by using the Eq. (\ref{eq:relation}) and (\ref{eq:mtof2}). Finally, we can express the efficiency corrected cumulants of net-proton distribution with 
the efficiency corrected moments by using the recursion relation:
\begin{equation} \label{eq:cumulants}
{C _r}({N_p} - {N_{\bar p}}) = {m_r}({N_p} - {N_{\bar p}}) - \sum\limits_{s = 1}^{r - 1} {\left( \begin{array}{c}
r - 1\\
s - 1
\end{array} \right)} {C _s}({N_p} - {N_{\bar p}}){m_{r - s}}({N_p} - {N_{\bar p}})
\end{equation}
where the $C_r$ denotes the $r^{th}$ order cumulants of net-proton distributions. 

In principle, one can also express the factorial moments in Eq. (\ref{eq:mtof2}) in terms of the cumulants and the various order efficiency corrected cumulants can be expressed by the measured cumulants and efficiency as :
\begin{equation} \label{eq:effcorrPRL}
\begin{split}
C_1^{X-Y}&=\frac{<x>-<y>}{\varepsilon} \\
C_2^{X - Y} &= \frac{{C_2^{x - y} + (\varepsilon  - 1)( < x >  +  < y > )}}{{{\varepsilon ^2}}}\\
C_3^{X - Y} &= \frac{{C_3^{x - y} + 3(\varepsilon  - 1)(C_2^x - C_2^y) + (\varepsilon  - 1)(\varepsilon  - 2)( < x >  -  < y > )}}{{{\varepsilon ^3}}}\\
C_4^{X - Y} &= \frac{{C_4^{x - y} - 2(\varepsilon  - 1)C_3^{x+y} + 8(\varepsilon  - 1)(C_3^x + C_3^y) + (5 - \varepsilon )(\varepsilon  - 1)C_2^{x + y}}}{{{\varepsilon ^4}}}\\
 &+ \frac{{8(\varepsilon  - 1)(\varepsilon  - 2)(C_2^x + C_2^y) + ({\varepsilon ^2} - 6\varepsilon  + 6)(\varepsilon  - 1)( < x >  +  < y > )}}{{{\varepsilon ^4}}}
\end{split}
\end{equation}
where the $(X,Y)$ and $(x,y)$ are the numbers of $(p, \bar{p})$ produced and measured, respectively. $\varepsilon=\varepsilon_p=\varepsilon_{\bar{p}}$ is the $p(\bar{p})$ efficiency. 
Obviously, the efficiency corrected cumulants are sensitive to the efficiency and depend on the lower order measured cumulants. 

In the previous discussion,  the detection efficiency of proton and anti-proton are considered to be constant within the entire phase space. In many cases, the efficiency of proton and anti-proton will depend on the phase space (transverse momentum ($p_{T}$), rapidity (y), azimuthal angle ($\phi$)). In this sense, one has to re-consider  the efficiency correction method.  In the paper~\cite{voker_eff2}, a new method for dealing with this case has been discussed, but the formulae for efficiency correction are rather involved and difficult to understand.  In the following, we will provide an alternative efficiency correction method for the phase space dependent efficiency, which is straightforward and easier to understand.  For simplify, we only consider the phase space of the proton and anti-proton are decomposed into two sub-phase spaces (1 and 2), within which the efficiency of proton and anti-proton are constant. We use the symbol $\varepsilon _{{p_1}}^{},\varepsilon _{{p_2}}^{}$
and $\varepsilon _{{{\bar p}_1}}^{},\varepsilon _{{{\bar p}_2}}^{}$ to denote the efficiency of proton and anti-proton in the two sub-phase spaces, and the corresponding number of proton and anti-proton in the two sub-phase spaces are $N_{p_1}$, $N_{p_2}$ and $N_{\bar{p}_1}$,  $N_{\bar{p}_2}$, respectively. Using the relations in Eq. (\ref{eq:mtof}) and (\ref{eq:ftom}), one has:

\begin{equation} \label{eq:4Fto2F}
\begin{split}
&{F_{{r_1},{r_2}}}({N_p},{N_{\bar p}}) = {F_{{r_1},{r_2}}}({N_{{p_1}}} + {N_{{p_2}}},{N_{{{\bar p}_1}}} + {N_{{{\bar p}_2}}})\\
 &= \sum\limits_{{i_1} = 0}^{{r_1}} {\sum\limits_{{i_2} = 0}^{{r_2}} {{s_1}({r_1},{i_1})} } {s_1}({r_2},{i_2})<{({N_{{p_1}}} + {N_{{p_2}}})^{{i_1}}}{({N_{{{\bar p}_1}}} + {N_{{{\bar p}_2}}})^{{i_2}}}>\\
& = \sum\limits_{{i_1} = 0}^{{r_1}} {\sum\limits_{{i_2} = 0}^{{r_2}} {{s_1}({r_1},{i_1})} } {s_1}({r_2},{i_2})< {\sum\limits_{s = 0}^{{i_1}} {\left( {\begin{array}{*{20}{c}}
{{i_1}}\\
s
\end{array}} \right)N_{{p_1}}^{{i_1} - s}N_{{p_2}}^s\sum\limits_{t = 0}^{{i_2}} {\left( {\begin{array}{*{20}{c}}
{{i_2}}\\
t
\end{array}} \right)N_{{{\bar p}_1}}^{{i_2} - t}N_{{{\bar p}_2}}^t} } } >\\
& = \sum\limits_{{i_1} = 0}^{{r_1}} {\sum\limits_{{i_2} = 0}^{{r_2}} {\sum\limits_{s = 0}^{{i_1}} {\sum\limits_{t = 0}^{{i_2}} {{s_1}({r_1},{i_1}){s_1}({r_2},{i_2})\left( {\begin{array}{*{20}{c}}
{{i_1}}\\
s
\end{array}} \right)\left( {\begin{array}{*{20}{c}}
{{i_2}}\\
t
\end{array}} \right)} } } } <N_{{p_1}}^{{i_1} - s}N_{{p_2}}^sN_{{{\bar p}_1}}^{{i_2} - t}N_{{{\bar p}_2}}^t>\\
&= \sum\limits_{{i_1} = 0}^{{r_1}} {\sum\limits_{{i_2} = 0}^{{r_2}} {\sum\limits_{s = 0}^{{i_1}} {\sum\limits_{t = 0}^{{i_2}} {\sum\limits_{u = 0}^{{i_1} - s} {\sum\limits_{v = 0}^s {\sum\limits_{j = 0}^{{i_2} - t} {\sum\limits_{k = 0}^t {{s_1}({r_1},{i_1}){s_1}({r_2},{i_2})\left( {\begin{array}{*{20}{c}}
{{i_1}}\\
s
\end{array}} \right)\left( {\begin{array}{*{20}{c}}
{{i_2}}\\
t
\end{array}} \right)} } } } } } } } \\
 &\times {s_2}({i_1} - s,u){s_2}(s,v){s_2}({i_2} - t,j){s_2}(t,k) \times {F_{u,v,j,k}}(N_{{p_1}}^{},N_{{p_2}}^{},N_{{{\bar p}_1}}^{},N_{{{\bar p}_2}}^{})
\end{split}
\end{equation}

Based on the Eq. (\ref{eq:4Fto2F}), we made a connection between the bivariate factorial moments of proton and anti-proton distributions in the entire phase space and
the multivariate factorial moments of proton and anti-proton distributions in the two sub-phase spaces. As a direct extension of Eq. (\ref{eq:relation}) for multivariate case, the efficiency corrected multivariate factorial moments of proton and anti-proton distributions in the sub-phase spaces can be obtained as:
 \begin{equation} \label{eq:relation2}
{F_{u,v,j,k}}(N_{{p_1}}^{},N_{{p_2}}^{},N_{{{\bar p}_1}}^{},N_{{{\bar p}_2}}^{}) = \frac{{{f_{u,v,j,k}}(n_{{p_1}}^{},n_{{p_2}}^{},n_{{{\bar p}_1}}^{},n_{{{\bar p}_2}}^{})}}{{{{({\varepsilon _{{p_1}}})}^u}{{({\varepsilon _{{p_2}}})}^v}{{({\varepsilon _{{{\bar p}_1}}})}^j}{{({\varepsilon _{{{\bar p}_2}}})}^k}}}\end{equation}
where  ${{f_{u,v,j,k}}(N_{{p_1}}^{},N_{{p_2}}^{},N_{{{\bar p}_1}}^{},N_{{{\bar p}_2}}^{})}$ is the measured multivariate factorial moments of proton and anti-proton distributions. If proton and anti-proton in the sub-phase space are independent, one has: 
 \begin{equation}{F_{u,v,j,k}}(N_{{p_1}}^{},N_{{p_2}}^{},N_{{{\bar p}_1}}^{},N_{{{\bar p}_2}}^{}) = {F_u}(N_{{p_1}}^{}){F_v}(N_{{p_2}}^{}){F_j}(N_{{{\bar p}_1}}^{}){F_k}(N_{{{\bar p}_2}}^{}) \end{equation}
By using  Eq. (\ref{eq:mtof2}), (\ref{eq:cumulants}), (\ref{eq:4Fto2F}) and (\ref{eq:relation2}), one can obtain the efficiency corrected moments of net-proton distributions for the case, where
the proton (anti-proton) are with different efficiency in two sub-phase spaces. If the efficiency of the proton (anti-proton) have large variations within the phase space, one needs to further divide the phase space into small ones. It is straightforward to do this, but it requires more computing resources for efficiency correction.
 
 \section{Error Estimation for Efficiency Corrected Moments} 
In the paper~\cite{Delta_theory}, we have already derived the error formulae for various order moments based on the Delta theorem in statistics. However, those formulae can only be applied to a special case, in which the efficiency of moments is assumed to be unity  ($\varepsilon=1$).  In the following, general error formulae 
for evaluating the statistical errors of efficiency corrected moments of conserved quantities in heavy-ion collisions will be derived based on the Delta theorem in statistics.  With those analytical formulae,  one can predict the expected errors with the number of events and efficiency numbers.
 \subsection{Delta theorem in Statsitics}
The Delta theorem is a well known theorem in statistics, which is used to approximate the distribution of a transformation
of a statistic in large samples if we can approximate the distribution of the statistic itself. Distributions of transformations of a statistic
are of great importance in applications. We will give the theorem without proofs and one can see~\cite{asytheory, junshan}.

{ \bf {\textit {Delta Theorem}}:} Suppose that
${\bf{ X}}=\{X_{1},X_{2},...,X_{k}\}$ is normally distributed as
$N({\bf{\mu}}, {\bf{\Sigma}}/n)$,  with $\bf \Sigma$ a covariance
matrix. Let ${\bf {g(x)}}=( g_{1}({\bf x}),...,g_{m}({\bf x}))$, ${\bf
{x}}=(x_{1},...x_{k})$, be a vector-valued function for which each
component function $g_{i}({\bf x})$ is real-valued and has a non-zero
differential $g_{i}(\mu)$, at ${\bf{x}}={\bf{\mu}}$. Put
\begin{equation}{\bf{D}} = \left[ {\left. {\frac{{\partial g_i
}}{{\partial x_j }}} \right|_{x = \mu } } \right]_{m \times
k}\end{equation} Then
\begin{equation} \label{eq:limitvar} {\bf{g}}({\bf{X}})\xrightarrow[]{d}N({\bf{g}}(\mu),\frac{{\bf{D\Sigma D^{'}}}}{n})\end{equation}
where $n$ is the number of events.

 \subsection{General Error Formula for Statistic Quantities}
Based on the Delta theorem, one can derive the general error formula for a statistic quantity. Suppose,  statistic quantity $\phi$ is as a function of random variables ${\bf{ X}}=\{X_{1},X_{2},...,X_{m}\}$, then the transformation functions ${\bf {g(X)}}=\phi({\bf X})$.  The {\bf D} matrix can be written as: 
\begin{equation}
{\bf{D}} = {\left[ {\frac{{\partial \phi }}{{\partial {\bf X}}}} \right]_{1\times m}}
\end{equation} 
and the covariance matrix $\Sigma$ is:
\begin{equation}
\Sigma  = n \times Cov({X_i},{X_j})
\end{equation} 
Based on Eq. (\ref{eq:limitvar}), the variance of the statistic $\phi$ can be calculated as:
\begin{equation} \label{eq:error}
\begin{split}
V(\phi ) &=\frac{{\bf{D\Sigma D^{'}}}}{n}= \sum\limits_{i = 1,j = 1}^m {\left( {\frac{{\partial \phi }}{{\partial {X_i}}}} \right)} \left( {\frac{{\partial \phi }}{{\partial {X_j}}}} \right)Cov({X_i},{X_j})\\
 &= \sum\limits_{i = 1}^m {{{\left( {\frac{{\partial \phi }}{{\partial {X_i}}}} \right)}^2}} V({X_i}) + \sum\limits_{i = 1,j = 1,i \ne j}^m {\left( {\frac{{\partial \phi }}{{\partial {X_i}}}} \right)} \left( {\frac{{\partial \phi }}{{\partial {X_j}}}} \right)Cov({X_i},{X_j})
\end{split}
\end{equation}
where $V(X_i)$ is the variance of variable $X_i$ and $Cov(X_i,X_j)$ is the covariance between $X_i$ and $X_j$. 
To calculate the statistical errors, one needs to know the variance and covariance of the variable $X_i$ and $X_j$ in the Eq. (\ref{eq:error}).  Since the efficiency corrected moments are expressed in terms of the factorial moments, the factorial moments are the random variable $X_i$ in Eq.  (\ref{eq:error}).  Then, we need to know the expression for variance and covariance of the factorial moments. It is known that the covariance of the multivariate moments~\cite{advancetheory} can be written as:
\begin{equation} \label{eq:covm}Cov({m_{r,s}},{m_{u,v}}) = \frac{1}{n}({m_{r + u,s + v}} - {m_{r,s}}{m_{u,v}}) \end{equation}
where $n$ is the number of events, $m_{r,s}=<X_1^{r}X_2^{s}>$ and ${m_{u,v}}=<X_1^{u}X_2^{v}>$ are the multivariate moments, the $X_1$ and $X_2$ are random variables. Based on Eq. (\ref{eq:ftom}) and (\ref{eq:covm}), one can obtain the 
covariance for the multivariate factorial moments as:
\begin{equation} \label{eq:covF}
\begin{split}
&Cov({f_{r,s}},{f_{u,v}}) = Cov\left(\sum\limits_{i = 0}^r {\sum\limits_{j = 0}^s {{s_1}(r,i){s_1}(s,j){m_{i,j}},} } \sum\limits_{k = 0}^u {\sum\limits_{h = 0}^v {{s_1}(u,k){s_1}(v,h){m_{k,h}}} } \right)\\
 &= \sum\limits_{i = 0}^r {\sum\limits_{j = 0}^s {\sum\limits_{k = 0}^u {\sum\limits_{h = 0}^v {{s_1}(r,i){s_1}(s,j){s_1}(u,k){s_1}(v,h)} }  \times Cov({m_{i,j}},{m_{k,h}})} } \\
 &= \frac{1}{n}\sum\limits_{i = 0}^r {\sum\limits_{j = 0}^s {\sum\limits_{k = 0}^u {\sum\limits_{h = 0}^v {{s_1}(r,i){s_1}(s,j){s_1}(u,k){s_1}(v,h)} }  \times } } ({m_{i + k,j + h}} - {m_{i,j}}{m_{k,h}})\\
&=\frac{1}{n}{\sum\limits_{i = 0}^r {\sum\limits_{j = 0}^s {\sum\limits_{k = 0}^u {\sum\limits_{h = 0}^v {\sum\limits_{\alpha  = 0}^{i + k} {\sum\limits_{\beta  = 0}^{j + h} {{s_1}(r,i){s_1}(s,j){s_1}(u,k){s_1}(v,h){s_2}(i + k,\alpha ){s_2}(j + h,\beta ){f_{\alpha ,\beta }}} } } } } } }  \\
&-\frac{1}{n}{f_{r,s}}{f_{u,v}} \\
 &= \frac{1}{n}({f_{(r,u),(s,v)}} - {f_{r,s}}{f_{u,v}})
\end{split}
\end{equation}
where the $f_{(r,u),(s,v)}$ is defined as:
\begin{equation} \label{eq:fact4}
\begin{split}
&{f_{(r,u),(s,v)}} = \left\langle {\frac{{{X_1}!}}{{({X_1} - r)!}}\frac{{{X_1}!}}{{({X_1} - u)!}}\frac{{{X_2}!}}{{({X_2} - s)!}}\frac{{{X_2}!}}{{({X_2} - v)!}}} \right\rangle \\
&={\sum\limits_{i = 0}^r {\sum\limits_{j = 0}^s {\sum\limits_{k = 0}^u {\sum\limits_{h = 0}^v {\sum\limits_{\alpha  = 0}^{i + k} {\sum\limits_{\beta  = 0}^{j + h} {{s_1}(r,i){s_1}(s,j){s_1}(u,k){s_1}(v,h){s_2}(i + k,\alpha ){s_2}(j + h,\beta ){f_{\alpha ,\beta }}} } } } } } }  
\end{split}
\end{equation}
The definition of bivariate factorial moments $f_{r,s}$, $f_{u,v}$ and $f_{\alpha,\beta}$ are the same as Eq. (\ref{eq:fact2}).
The Eq. (\ref{eq:covF}) can be put into the standard error propagation formulae (\ref{eq:error}) to calculate the statistical errors of the efficiency corrected moments.

For simplify, we assume the efficiency of proton and anti-proton is constant within the entire phase space. The original and measured numbers of proton and anti-proton are denoted as $N_p$, $N_{\bar{p}}$ and $n_p$ , $n_{\bar{p}}$ ,  the efficiency of proton and anti-proton are $\varepsilon_p$ and $\varepsilon_{\bar{p}}$, respectively. We also define the symbols for moments, cumulants and factorial moments of proton and anti-proton as:
\begin{enumerate}
\item {\bf Statistic Quantity of Net-proton Distribution}: $\Phi ({N_p} - {N_{\bar p}})$. \\
It denotes any efficiency corrected statistic quantities of net-proton distribution, such as moments, cumulants and ratio of cumulants.
\item {\bf  Central Moments of Net-proton Distribution }: $\mu_{r}$. \\
This symbol represents $r^{th}$ order efficiency corrected central moments of net-proton distribution.
 $$\mu_{r}=<(N_p - N_{\bar p}-<N_p - N_{\bar p}>)^{r}>$$
\item {\bf  Joint Moments of Proton and Anti-proton Distribution }: $m_{s,t}$ \\
It denotes efficiency corrected $s^{th}$ and $t^{th}$  order joint moments of proton and anti-proton distribution, respectively.
$$m_{s,t}=<N_p^sN_{\bar p}^t>$$
\item {\bf  Joint Factorial Moments of Proton and Anti-proton Distribution }: $F_{i,j}$ and $f_{i,j}$ \\
The symbol $F_{i,j}$ denotes $i^{th}$ and $j^{th}$ order efficiency corrected joint factorial moments of proton and anti-proton distributions, respectively.
$f_{i,j}$ denotes $i^{th}$ and $j^{th}$  order efficiency uncorrected joint factorial moments of proton and anti-proton distributions, respectively.
One can see the definitions for $F_{i,j}$  and $f_{i,j}$  in the Eq. (\ref{eq:fact1}) and (\ref{eq:fact2}), respectively.
\end{enumerate}

With those definitions, one can write down the variance of the efficiency corrected statistic quantity of net-proton distributions based on standard error propagation:
\begin{equation} \label{eq:errorp1}
\begin{split}
&V(\Phi ({N_p} - {N_{\bar p}})) \\
&= \sum\limits_{r = 1}^H {\sum\limits_{s,t = 0}^H {\sum\limits_{i,j = 0}^H {\sum\limits_{u,v = 0}^H {\left( {\frac{{\partial \Phi }}{{\partial {\mu _r}}}\frac{{\partial {\mu _r}}}{{\partial {m_{s,t}}}}\frac{{\partial {m_{s,t}}}}{{\partial F_{i,j}^{}}}\frac{{\partial F_{i,j}^{}}}{{\partial f_{i,j}^{}}}} \right)} \left( {\frac{{\partial \Phi }}{{\partial {\mu _r}}}\frac{{\partial {\mu _r}}}{{\partial {m_{s,t}}}}\frac{{\partial {m_{s,t}}}}{{\partial F_{u,v}^{}}}\frac{{\partial F_{u,v}^{}}}{{\partial f_{u,v}^{}}}} \right)} } Cov({f_{i,j}},{f_{u,v}})} \\
 &= \frac{1}{n}\sum\limits_{i,j = 0}^H {\sum\limits_{u,v = 0}^H {\frac{{{D_{i,j}}{D_{u,v}}}}{{\varepsilon _p^{i + u}\varepsilon _{\bar p}^{j + v}}}} } Cov({f_{i,j}},{f_{u,v}})\\
 &= \frac{1}{n}\sum\limits_{i,j = 0}^H {\sum\limits_{u,v = 0}^H {\frac{{{D_{i,j}}{D_{u,v}}}}{{\varepsilon _p^{i + u}\varepsilon _{\bar p}^{j + v}}}} } ({f_{(i,u),(j,v)}} - {f_{i,j}}{f_{u,v}})
\end{split}
\end{equation}
where $n$ is the number of events,  $H$ is the highest power of central moments in the statistic quantity,  $Cov({f_{i,j}},{f_{u,v}})$ is  the covariance between efficiency uncorrected factorial moments and can be calculated via Eq. (\ref{eq:fact4}), $D_{i,j}$ and $D_{u,v}$ are efficiency corrected differential coefficients and are calculated as :
\begin{equation} \label{eq:differential}
{D_{i,j}} = \sum\limits_{r = 1}^H {\sum\limits_{s,t = 0}^H {\left( {\frac{{\partial \Phi }}{{\partial \mu_r}}\frac{{\partial \mu _r}}{{\partial {m_{s,t}}}}\frac{{\partial {m_{s,t}}}}{{\partial F_{i,j}^{}}}} \right)} },{D_{u,v}} = \sum\limits_{r = 1}^H {\sum\limits_{s,t = 0}^H {\left( {\frac{{\partial \Phi }}{{\partial \mu _r}}\frac{{\partial \mu _r}}{{\partial {m_{s,t}}}}\frac{{\partial {m_{s,t}}}}{{\partial F_{u,v}^{}}}} \right)} } \end{equation}
In the Eq. (\ref{eq:errorp1}), the derivation of the efficiency uncorrected factorial moments is decomposed into several steps and the efficiency numbers will appear as the denominator in the formula in the final derivation step. 

If the efficiency of proton (anti-proton) is not constant within the entire phase space, it is straightforward  to update the Eq. (\ref{eq:errorp1}). For simplify, we assume that the phase space can be divided into 2 sub-phase spaces, within which the efficiency of proton (anti-proton) is constant. The same sets of notation as used in the section 2 are used for this discussion. As a direct extension, one can re-write the Eq. (\ref{eq:covm}) and (\ref{eq:covF}) for multivariate case.   
\begin{align}
&Cov({m_{r,s,u,v}},{m_{i,j,k,h}}) = \frac{1}{n}({m_{r + i,s + j,u + k,v + h}} - {m_{r,s,u,v}}{m_{i,j,k,h}})\\
&Cov({f_{r,s,u,v}},{f_{i,j,k,h}}) = \frac{1}{n}({f_{(r,i),(s,j),(u,k),(v,h)}} - {f_{r,s,u,v}}{f_{i,j,k,h}})
\end{align}
For random variable $X_1$, $X_2$, $X_3$ and $X_4$, the ${f_{(r,i),(s,j),(u,k),(v,h)}}$ is defined as:
\begin{equation} \label{eq:factcov4}
\begin{split}
&{f_{(r,i),(s,j),(u,k),(v,h)}} = \left\langle {\frac{{{X_1}!}}{{({X_1} - r)!}}\frac{{{X_1}!}}{{({X_1} - i)!}}\frac{{{X_2}!}}{{({X_2} - s)!}}\frac{{{X_2}!}}{{({X_2} - j)!}}} \right.\\
& \times \left. {\frac{{{X_3}!}}{{({X_3} - u)!}}\frac{{{X_3}!}}{{({X_3} - k)!}}\frac{{{X_4}!}}{{({X_4} - v)!}}\frac{{{X_4}!}}{{({X_4} - h)!}}} \right\rangle \\
 &= \sum\limits_{{x_1} = 0}^r {\sum\limits_{{x_2} = 0}^s {\sum\limits_{{x_3} = 0}^u {\sum\limits_{{x_4} = 0}^v {\sum\limits_{{y_1} = 0}^i {\sum\limits_{{y_2} = 0}^j {\sum\limits_{{y_3} = 0}^k {\sum\limits_{{y_4} = 0}^h {\sum\limits_{\alpha  = 0}^{{x_1} + {y_1}} {\sum\limits_{\beta  = 0}^{{x_2} + {y_2}} {\sum\limits_{\gamma  = 0}^{{x_3} + {y_3}} {\sum\limits_{\delta  = 0}^{{x_4} + {y_4}} {{s_1}(r,{x_1}){s_1}(s,,{x_2})} } } } } } } } } } } } \\
 &\times {s_1}(u,{x_3}){s_1}(v,{x_4}) \times {s_1}(i,{y_1}){s_1}(j,,{y_2}){s_1}(k,{y_3}){s_1}(h,{y_4})\\
 &\times {s_2}({x_1} + {y_1},\alpha ){s_2}({x_2} + {y_2},\beta ){s_2}({x_3} + {y_3},\gamma ){s_2}({x_4} + {y_4},\delta ){f_{\alpha ,\beta ,\gamma ,\delta }}
\end{split}
\end{equation}
Then, the Eq. (\ref{eq:errorp1})  can be re-written as:
\begin{equation} \label{eq:errorp2}
\begin{split}
&V(\Phi ({N_p} - {N_{\bar p}}))\\
 &= \sum\limits_{r = 1}^H {\sum\limits_{s,t = 0}^H {\sum\limits_{i,j = 0}^H {\sum\limits_{u,v = 0}^H {\sum\limits_{\scriptstyle{\alpha _1},{\alpha _2},{{\bar \alpha }_1},{{\bar \alpha }_2} = 0\hfill\atop
\scriptstyle{\beta _1},{\beta _2},{{\bar \beta }_1},{{\bar \beta }_2} = 0\hfill}^H {\left( {\frac{{\partial \Phi }}{{\partial {\mu _r}}}\frac{{\partial {\mu _r}}}{{\partial {m_{s,t}}}}\frac{{\partial {m_{s,t}}}}{{\partial F_{i,j}^{}}}\frac{{\partial F_{i,j}^{}}}{{\partial F_{{\alpha _1},{\alpha _2},{{\bar \alpha }_1},{{\bar \alpha }_2}}^{}}}\frac{{\partial F_{{\alpha _1},{\alpha _2},{{\bar \alpha }_1},{{\bar \alpha }_2}}^{}}}{{\partial f_{{\alpha _1},{\alpha _2},{{\bar \alpha }_1},{{\bar \alpha }_2}}^{}}}} \right)} } } } } \\
& \times \left( {\frac{{\partial \Phi }}{{\partial {\mu _r}}}\frac{{\partial {\mu _r}}}{{\partial {m_{s,t}}}}\frac{{\partial {m_{s,t}}}}{{\partial F_{u,v}^{}}}\frac{{\partial F_{u,v}^{}}}{{\partial F_{{\beta _1},{\beta _2},{{\bar \beta }_1},{{\bar \beta }_2}}^{}}}\frac{{\partial F_{{\beta _1},{\beta _2},{{\bar \beta }_1},{{\bar \beta }_2}}^{}}}{{\partial {f_{{\beta _1},{\beta _2},{{\bar \beta }_1},{{\bar \beta }_2}}}}}} \right)Cov(f_{{\alpha _1},{\alpha _2},{{\bar \alpha }_1},{{\bar \alpha }_2}}^{},f_{{\beta _1},{\beta _2},{{\bar \beta }_1},{{\bar \beta }_2}}^{})\\
& = \sum\limits_{\scriptstyle{\alpha _1},{\alpha _2},{{\bar \alpha }_1},{{\bar \alpha }_2} = 0\hfill\atop
\scriptstyle{\beta _1},{\beta _2},{{\bar \beta }_1},{{\bar \beta }_2} = 0\hfill}^H {\frac{{D_{{\alpha _1},{\alpha _2},{{\bar \alpha }_1},{{\bar \alpha }_2}}^{}D_{{\beta _1},{\beta _2},{{\bar \beta }_1},{{\bar \beta }_2}}^{}Cov(f_{{\alpha _1},{\alpha _2},{{\bar \alpha }_1},{{\bar \alpha }_2}}^{},f_{{\beta _1},{\beta _2},{{\bar \beta }_1},{{\bar \beta }_2}}^{})}}{{\varepsilon _{{p_1}}^{{\alpha _1} + {\beta _1}}\varepsilon _{{p_2}}^{{\alpha _2} + {\beta _2}}\varepsilon _{{{\bar p}_1}}^{{{\bar \alpha }_1} + {{\bar \beta }_1}}\varepsilon _{{{\bar p}_2}}^{{{\bar \alpha }_2} + {{\bar \beta }_2}}}}}   \\
& = \frac{1}{n}\sum\limits_{\scriptstyle{\alpha _1},{\alpha _2},{{\bar \alpha }_1},{{\bar \alpha }_2} = 0\hfill\atop
\scriptstyle{\beta _1},{\beta _2},{{\bar \beta }_1},{{\bar \beta }_2} = 0\hfill}^H {\left[ {\frac{{D_{{\alpha _1},{\alpha _2},{{\bar \alpha }_1},{{\bar \alpha }_2}}^{}D_{{\beta _1},{\beta _2},{{\bar \beta }_1},{{\bar \beta }_2}}^{}}}{{\varepsilon _{{p_1}}^{{\alpha _1} + {\beta _1}}\varepsilon _{{p_2}}^{{\alpha _2} + {\beta _2}}\varepsilon _{{{\bar p}_1}}^{{{\bar \alpha }_1} + {{\bar \beta }_1}}\varepsilon _{{{\bar p}_2}}^{{{\bar \alpha }_2} + {{\bar \beta }_2}}}}} \right.}  
\left. {  ({f_{({\alpha _1},{\beta _1}),({\alpha _2},{\beta _2}),({{\bar \alpha }_1},{{\bar \beta }_1}),({{\bar \alpha }_2},{{\bar \beta }_2})}} - f_{{\alpha _1},{\alpha _2},{{\bar \alpha }_1},{{\bar \alpha }_2}}^{}f_{{\beta _1},{\beta _2},{{\bar \beta }_1},{{\bar \beta }_2}}^{})} \right]
\end{split}
\end{equation}
where $n$ is the number of events,  $H$ is the highest power of central moments in the statistic quantity $\Phi$, $D_{{\alpha _1},{\alpha _2},{{\bar \alpha }_1},{{\bar \alpha }_2}}^{}$ and $D_{{\beta _1},{\beta _2},{{\bar \beta }_1},{{\bar \beta }_2}}^{}$  are defined as:
\begin{equation}
D_{{\alpha _1},{\alpha _2},{{\bar \alpha }_1},{{\bar \alpha }_2}}^{} = \sum\limits_{r = 1}^H {\sum\limits_{s,t = 0}^H {\sum\limits_{i,j = 0}^H {\frac{{\partial \Phi }}{{\partial {\mu _r}}}\frac{{\partial {\mu _r}}}{{\partial {m_{s,t}}}}\frac{{\partial {m_{s,t}}}}{{\partial F_{i,j}^{}}}\frac{{\partial F_{i,j}^{}}}{{\partial F_{{\alpha _1},{\alpha _2},{{\bar \alpha }_1},{{\bar \alpha }_2}}^{}}}} }} 
\end{equation}
\begin{equation}
D_{{\beta _1},{\beta _2},{{\bar \beta }_1},{{\bar \beta }_2}}^{} = \sum\limits_{r = 1}^H {\sum\limits_{s,t = 0}^H {\sum\limits_{u,v= 0}^H {\frac{{\partial \Phi }}{{\partial {\mu _r}}}\frac{{\partial {\mu _r}}}{{\partial {m_{s,t}}}}\frac{{\partial {m_{s,t}}}}{{\partial F_{i,j}^{}}}\frac{{\partial F_{i,j}^{}}}{{\partial F_{{\beta _1},{\beta _2},{{\bar \beta }_1},{{\bar \beta }_2}}^{}}}} }} 
\end{equation}

The $F_{{\alpha _1},{\alpha _2},{{\bar \alpha }_1},{{\bar \alpha }_2}}^{}$ and $f_{{\alpha _1},{\alpha _2},{{\bar \alpha }_1},{{\bar \alpha }_2}}^{}$ are definied as:
\begin{equation}
F_{{\alpha _1},{\alpha _2},{{\bar \alpha }_1},{{\bar \alpha }_2}}^{} = \left\langle {\frac{{{N_{{p_1}}}!}}{{({N_{{p_1}}} - {\alpha _1})!}}\frac{{{N_{{p_2}}}!}}{{({N_{{p_2}}} - {\alpha _2})!}}\frac{{{N_{{{\bar p}_1}}}!}}{{({N_{{{\bar p}_1}}} - {{\bar \alpha }_1})!}}\frac{{{N_{{{\bar p}_2}}}!}}{{({N_{{{\bar p}_2}}} - {{\bar \alpha }_2})!}}} \right\rangle 
\end{equation}
\begin{equation}
f_{{\alpha _1},{\alpha _2},{{\bar \alpha }_1},{{\bar \alpha }_2}}^{} = \left\langle {\frac{{{n_{{p_1}}}!}}{{({n_{{p_1}}} - {\alpha _1})!}}\frac{{{n_{{p_2}}}!}}{{({n_{{p_2}}} - {\alpha _2})!}}\frac{{{n_{{{\bar p}_1}}}!}}{{({n_{{{\bar p}_1}}} - {{\bar \alpha }_1})!}}\frac{{{n_{{{\bar p}_2}}}!}}{{({n_{{{\bar p}_2}}} - {{\bar \alpha }_2})!}}} \right\rangle 
\end{equation}
where $N_{p_1}$, $N_{p_2}$, $N_{\bar p_1}$ and $N_{\bar p_2}$ ($N_{p}$=$N_{p_1}$+$N_{p_2}$, $N_{\bar p}$=$N_{\bar p_1}$+$N_{\bar p_2}$) are original number of proton and anti-proton in two sub-phase spaces (1, 2), respectively, $n_{p_1}$, $n_{p_2}$, $n_{\bar p_1}$ and $n_{\bar p_2}$ are the corresponding measured ones,  ${\varepsilon_{p_1}}$, ${\varepsilon_{p_2}}$, ${\varepsilon_{\bar p_1}}$ and  ${\varepsilon_{\bar p_2}}$ are the efficiency of proton and anti-proton in the two different sub-phase spaces, respectively. The relation between $F_{i,j}$ and $F_{{\alpha _1},{\alpha _2},{{\bar \alpha }_1},{{\bar \alpha }_2}}^{}$  is presented in the Eq. (\ref{eq:4Fto2F}). To calculate $F_{u,v}$ and $F_{{\beta _1},{\beta_2},{{\bar \beta }_1},{{\bar \beta }_2}}^{}$, one needs to replace the subscripts. By using standard error propagation, we express the variance of a statistic quantity in terms of factorial moments and efficiency of proton and anti-proton. In principle, once having the factorial moments and efficiency number, one can calculate the errors of statistic quantities, which can be decomposed into factorial moments.

\subsection{Derivation of Error Formula for Cumulants with Constant Efficiency }
In this section, for illustration purpose,  we only derive the general error formula for the efficiency corrected mean and variance of net-proton distributions. The efficiency of proton and anti-proton is assumed to be constant within entire phase space. When the efficiency numbers are set to 100\% for both proton and anti-proton, the error formulae will be reduced to the ones presented in the paper~\cite{Delta_theory}. As the expression for the error formulae of efficiency corrected higher order moments are rather involved, we will present those results in the appendix. 

\subsubsection{Mean (M)}

The efficiency corrected mean values of net-proton distributions can be calculated as:
\begin{equation} \label{eq:mean}	M = <{N_p}> -< {N_{\bar p}}> = <\frac{{{n_p}}}{{{\varepsilon _p}}} >-< \frac{{{n_{\bar p}}}}{{{\varepsilon _{\bar p}}}}> \end{equation}
The variance of the mean value can be calculated as:
\begin{equation} \label{eq:mean}
\begin{split}
V(M)&= {\left( {\frac{{\partial M}}{{\partial  < {N_p} > }}\frac{{\partial  < {N_p} > }}{{\partial  < {n_p} > }}} \right)^2}V( < {n_p} > ) + {\left( {\frac{{\partial M}}{{\partial  < {N_{\bar p}} > }}\frac{{\partial  < {N_{\bar p}} > }}{{\partial  < {n_{\bar p}} > }}} \right)^2}V( < {n_{\bar p}} > )\\
 &- 2\left( {\frac{{\partial M}}{{\partial  < {N_p} > }}\frac{{\partial  < {N_p} > }}{{\partial  < {n_p} > }}} \right)\left( {\frac{{\partial M}}{{\partial  < {N_{\bar p}} > }}\frac{{\partial  < {N_{\bar p}} > }}{{\partial  < {n_{\bar p}} > }}} \right)Cov( < {n_p} > , < {n_{\bar p}} > )\\
 &= \frac{1}{{\varepsilon _p^2}}V( < {n_p} > ) + \frac{1}{{\varepsilon _{\bar p}^2}}V( < {n_{\bar p}} > ) - 2\frac{1}{{{\varepsilon _p}{\varepsilon _{\bar p}}}}Cov( < {n_p} > , < {n_{\bar p}} > )\\
 & = \frac{1}{n}\left[ {\frac{1}{{\varepsilon _p^2}}V({n_p}) + \frac{1}{{\varepsilon _{\bar p}^2}}V({n_{\bar p}}) - \frac{2}{{{\varepsilon _p}{\varepsilon _{\bar p}}}}Cov({n_p},{n_{\bar p}})} \right]
\end{split}
\end{equation}
where $n$ is the number of events, $Cov({n_p},{n_{\bar p}}) =  < {n_p}{n_{\bar p}} >  -  < {n_p} >  < {n_{\bar p}} > $ is the covariance between measured number of proton and anti-proton.
If the efficiency of proton and anti-proton are equal, $\varepsilon  = {\varepsilon _p} = {\varepsilon _{\bar p}}$, then the variance can be written as:
\begin{equation} V(M) = \frac{1}{{n{\varepsilon ^2}}}V({n_p} - {n_{\bar p}})\end{equation}
and the statistical error for the efficiency corrected mean number of net-proton is: 
\begin{equation}\sigma (M = <{N_p} - {N_{\bar p}}>) = \sqrt {V(M)}  = \frac{1}{\varepsilon }\frac{{\sqrt {V({n_p} - {n_{\bar p}})} }}{{\sqrt n }} = \frac{1}{\varepsilon }\frac{{\sigma ({n_p} - {n_{\bar p}})}}{{\sqrt n }}\end{equation}

\subsubsection{Variance($\sigma^{2}$)}
Firstly, one can express the efficiency corrected variance of net-proton distribution in terms of various order factorial moments of proton and anti-proton distributions.
\begin{equation}
{\sigma ^2}({N_p} - {N_{\bar p}}) = {F_{0,1}} + {F_{0,2}} + {F_{1,0}} - 2{F_{1,1}} + {F_{2,0}} - {({F_{1,0}} - {F_{0,1}})^2} 
\end{equation}
As the highest order is 2, the variance of the variance can be calculated as:
\begin{equation} \label{eq:varofvar}
\begin{split}
V({\sigma ^2}({N_p} - {N_{\bar p}})) &= \sum\limits_{i,j = 0}^2 {\sum\limits_{u,v = 0}^2 {\frac{1}{{\varepsilon _p^{i + u}\varepsilon _{\bar p}^{j + v}}}\frac{{\partial {\sigma ^2}}}{{\partial {F_{i,j}}}}\frac{{\partial {\sigma ^2}}}{{\partial {F_{u,v}}}}} } Cov({f_{i,j}},{f_{u,v}})\\
 &= \sum\limits_{i,j = 0}^2 {\sum\limits_{u,v = 0}^2 {\frac{1}{{\varepsilon _p^{i + u}\varepsilon _{\bar p}^{j + v}}}{D_{i,j}}{D_{u,v}}} } Cov({f_{i,j}},{f_{u,v}})
\end{split}
\end{equation}
The basic variables are $F_{2,0}$, $F_{1,1}$, $F_{0,2}$, $F_{1,0}$ and $F_{0,1}$, we have the differential coefficient $D$ as:
\begin{equation}
{D_{2,0}} = 1,{D_{1,1}} =  - 2,{D_{0,2}} = 1,{D_{1,0}} = 1 - 2M,{D_{0,1}} = 1 + 2M 
\end{equation}
where $M=F_{1,0}- F_{0,1}$ is the efficiency corrected mean value of net-proton distribution, $Cov({f_{i,j}},{f_{u,v}})$ is the covariance between efficiency uncorrected 
joint factorial moments of proton and anti-proton, which can be calculated by using Eq. (\ref{eq:covF}) and (\ref{eq:fact4}). 
\begin{equation}
\begin{split}
&Cov({f_{r,s}},{f_{u,v}})  =\\
 &\frac{1}{n}\left( {\sum\limits_{i = 0}^r {\sum\limits_{j = 0}^s {\sum\limits_{k = 0}^u {\sum\limits_{h = 0}^v {\sum\limits_{\alpha  = 0}^{i + k} {\sum\limits_{\beta  = 0}^{j + h} {{s_1}(r,i){s_1}(s,j){s_1}(u,k){s_1}(v,h){s_2}(i + k,\alpha ){s_2}(j + h,\beta ){f_{\alpha ,\beta }}} } } } } }  - {f_{r,s}},{f_{u,v}}} \right)
 \end{split}
\end{equation}
Where $f_{r,s}$ andf $f_{u,v}$ are the measured joint factorial moments of proton and anti-proton. For simplify,  only several results of covariance are listed:  
\begin{equation}
\begin{split}
Cov({f_{2,0}},{f_{2,0}}) &= V({f_{2,0}}) =(2{f_{2,0}} - f_{2,0}^2 + 4{f_{3,0}} + {f_{4,0}})/n\\
Cov({f_{1,1}},{f_{1,1}}) &= V({f_{1,1}}) = ({f_{1,1}} - f_{1,1}^2 + {f_{1,2}} + {f_{2,1}} + {f_{2,2}})/n\\
Cov({f_{1,0}},{f_{1,0}}) &= V({f_{1,0}}) = ({f_{1,0}} - f_{1,0}^2 + {f_{2,0}})/n\\
Cov({f_{1,1}},{f_{2,0}}) &= Cov({f_{2,0}},{f_{1,1}}) =  (- {f_{1,1}}{f_{2,0}} + 2{f_{2,1}} + {f_{3,1}})/n\\
Cov({f_{1,0}},{f_{1,1}}) &= Cov({f_{1,1}},{f_{1,0}}) = ({f_{1,1}} - {f_{1,0}}{f_{1,1}} + {f_{2,1}})/n\\
Cov({f_{0,2}},{f_{1,0}}) &= Cov({f_{1,0}},{f_{0,2}}) =  (- {f_{0,2}}{f_{1,0}} + {f_{1,2}})/n\\
Cov({f_{0,1}},{f_{1,0}}) &= Cov({f_{1,0}},{f_{0,1}}) = ( - {f_{0,1}}{f_{1,0}} + {f_{1,1}})/n
 \end{split}
\end{equation}
By assuming $\varepsilon=\varepsilon_p=\varepsilon_{\bar p} $, we can obtain the variance of the variance with Eq. (\ref{eq:varofvar}):
\begin{equation} 
V({\sigma ^2}({N_p} - {N_{\bar p}})) = \frac{1}{n}\left( {\frac{A}{{{\varepsilon ^4}}} + \frac{B}{{{\varepsilon ^3}}} + \frac{C}{{{\varepsilon ^2}}}} \right)
\end{equation}
and the statistical error of variance of net-proton distribution is:
\begin{equation} \label{eq:err_var}  
\sigma ({\sigma ^2}({N_p} - {N_{\bar p}})) = \frac{1}{{\sqrt n }}\sqrt {\left( {\frac{A}{{{\varepsilon ^4}}} + \frac{B}{{{\varepsilon ^3}}} + \frac{C}{{{\varepsilon ^2}}}} \right)} 
\end{equation}
where $n$ is the number of events, $A$, $B$ and $C$ are defined as:
\begin{equation}  
\begin{split}
A &=  2{f_{0,2}}- f_{0,2}^2 - f_{2,0}^2 + 4{f_{0,3}} + {f_{0,4}} - 4\left( { - {f_{0,2}}{f_{1,1}} + 2{f_{1,2}} + {f_{1,3}}} \right)\\
 &+ 2{f_{2,0}} + 2\left( {{f_{2,2}} - {f_{0,2}}{f_{2,0}}} \right) + 4\left( { - f_{1,1}^2 + {f_{1,1}} + {f_{1,2}} + {f_{2,1}} + {f_{2,2}}} \right)\\
 &+ 4{f_{3,0}} - 4\left( { - {f_{1,1}}{f_{2,0}} + 2{f_{2,1}} + {f_{3,1}}} \right) + {f_{4,0}} \\
B &= 2(2M + 1)\left( { - {f_{0,1}}{f_{0,2}} + 2{f_{0,2}} + {f_{0,3}}} \right) + 2(1 - 2M)\left( {{f_{1,2}} - {f_{0,2}}{f_{1,0}}} \right)\\
 &- 4(2M + 1)\left( { - {f_{0,1}}{f_{1,1}} + {f_{1,1}} + {f_{1,2}}} \right) + 2(2M + 1)\left( {{f_{2,1}} - {f_{0,1}}{f_{2,0}}} \right)\\
 &- 4(1 - 2M)\left( { - {f_{1,0}}{f_{1,1}} + {f_{1,1}} + {f_{2,1}}} \right) + 2(1 - 2M)\left( { - {f_{1,0}}{f_{2,0}} + 2{f_{2,0}} + {f_{3,0}}} \right) \\
C &= {(2M + 1)^2}\left( { - f_{0,1}^2 + {f_{0,1}} + {f_{0,2}}} \right) + 2(1 - 2M)(2M + 1)\left( {{f_{1,1}} - {f_{0,1}}{f_{1,0}}} \right)\\
& + {(1 - 2M)^2}\left( { - f_{1,0}^2 + {f_{1,0}} + {f_{2,0}}} \right)
\nonumber
\end{split}
\end{equation}
where $\varepsilon$ is the efficiency of proton and anti-proton.  The  constant $A$, $B$ and $C$
only consist of measured factorial moments of proton and anti-proton distributions. If $\varepsilon=1$, the error formula should be equivalent to
the ones presented in the paper~\cite{Delta_theory}, thus we have: $A + B + C = {\mu _4} - \mu _2^2$, where
$\mu_2$  and $\mu_4$ are the $2^{nd}$  and $4^{th}$ order central moments of net-proton distributions.

\section{Monte Carlo Simulation}
In this section, tests for the Delta theorem and Bootstrap error estimation on the efficiency corrected moments are made by using Monte Carlo (MC) simulation. The MC simulation is based on a skellam distribution, which is the distribution of the difference between two independent Poisson distribution. The probability density function of skellam distribution can be written as:
\begin{equation}
f(k;\mu _1 ,\mu _2 ) = e^{ -
(\mu _1 {{ + }}\mu _{{2}} )} (\frac{{\mu _1 }}{{\mu _{{2}} }})^{k/2}
I_{|k|} (2\sqrt {\mu _1 \mu _2 } ) \end{equation}
where the input parameters $\mu_1$ and $\mu_2$
are the mean value of the two Poisson distributions, respectively, the
$I_{k}(z)$ is the modified bessel function of the first kind.  Here, we use the skellam distribution as approximation of the net-proton distributions in the 
heavy-ion collisions and the $\mu_1$ ($\mu_2$) is the mean number of proton (anti-proton). The cumulants of skellam distribution can be calculated as:
\begin{equation}
{C_{2n}} = {\mu _1} + {\mu _2},{C_{2n - 1}} = {\mu _1} - {\mu _2} ,   (n = 1,2,3, \cdots )\end{equation}
Where the $C_{2n-1}$ and $C_{2n}$ are odd and even order cumulants, respectively. The odd order cumulants are equal to the difference between the two input parameters while the even order are the sum of the two 
parameters. Then, the cumulant ratios $C_{4}/C_2$ and $C_3/C_2$ of skellam distribution can be expressed as:
\begin{equation}
\begin{split}
&\kappa {\sigma ^2} = \frac{{{C_4}}}{{{C_2}}} = 1\\
&S\sigma  = \frac{{{C_3}}}{{{C_2}}} = \frac{{{\mu _1} - {\mu _2}}}{{{\mu _1} + {\mu _2}}}
\end{split}
\end{equation}
where $\kappa$ and $S$ are kurtosis and skewness, respectively. In the simulation, the two input parameters of skellam distribution are set to ${\mu _1}=6$ and ${\mu _2}=3$, which are close to the mean value of proton and anti-proton number in 0-5\% central Au+Au collisions at {\sNN}= 200 GeV. The efficiency of proton and anti-proton is set to be equal, $\varepsilon_p=\varepsilon_{\bar p}$.  To obtain the distributions with efficiency effects, numbers of proton and ant-proton are randomly and independently sampled from the original skellam distribution and  meanwhile,  every proton or anti-proton is forced to take a Bernoulli yes or no trial with a certain efficiency number to determine to accept it or not. 
\begin{figure}[htbp]
\begin{center}
\includegraphics[scale=0.5]{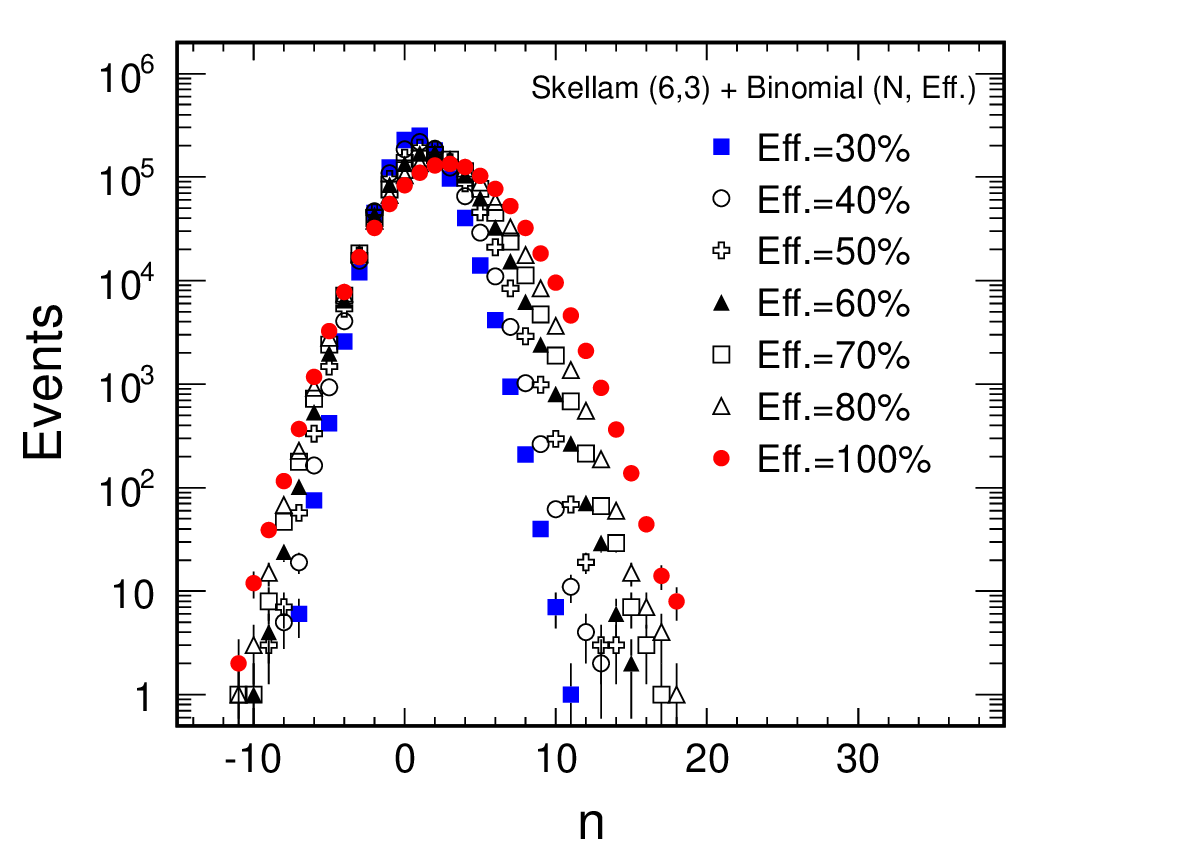}
\caption[]{(Color online) The event-by-event net-proton distributions obtained from the convolution between skellam distribution and binomial distribution with various efficiency parameters. 
The red solid circle represents the original skellam distribution without efficiency effects. The number of events is one million for each distribution. } \label{fig:dis}
\end{center}\end{figure}

Figure \ref{fig:dis} shows event-by-event net-proton distributions, which are constructed from the convolution between skellam distribution and binomial distribution with various efficiency numbers.  Seven cases have been studied with efficiency varying from 30\% to 100\%, with 10\% interval. The efficiency 100\% is the case without efficiency effects.  Fig. \ref{fig:dis} shows that when one decreases the efficiency, the mean value of the distribution become smaller and the shape of distributions is also different from the original one due to the efficiency effects. Based on the Eq. (\ref{eq:effcorrPRL}), the measured variance can be obtained via the mean and variance of the original distributions as:
\begin{equation} \label{eq:C2C2} C_2^{x - y} = C_2^{X - Y}[{\varepsilon ^2} + (1 - \varepsilon )\varepsilon \frac{{ < X >  +  < Y > }}{{C_2^{X - Y}}}]\end{equation}
For the skellam distribution, its variance is equal to the sum of the mean values of variable $X$ and $Y$, $C_2^{X-Y}=<X>+<Y>$,  then one has $C_2^{x - y} = \varepsilon C_2^{X - Y}$ from Eq. (\ref{eq:C2C2}). That's why the width of distributions with efficiency effects are narrower than the width of the original skellam distribution. Generally, it is not straightforward to tell whether the measured variance become larger or smaller without knowing the mean and variance of the original distributions. Similarly, one can also obtain $C_3^{x - y} = \varepsilon C_3^{X - Y}$ and ${S^{x - y}} = {S^{X - Y}}/\sqrt \varepsilon$ ($S$ is for skewness), which indicate the measured distributions are skewed comparing with the original skellam distribution.
\begin{figure}[htbp]
\begin{center}
\hspace{-1.5cm}
\includegraphics[scale=0.75]{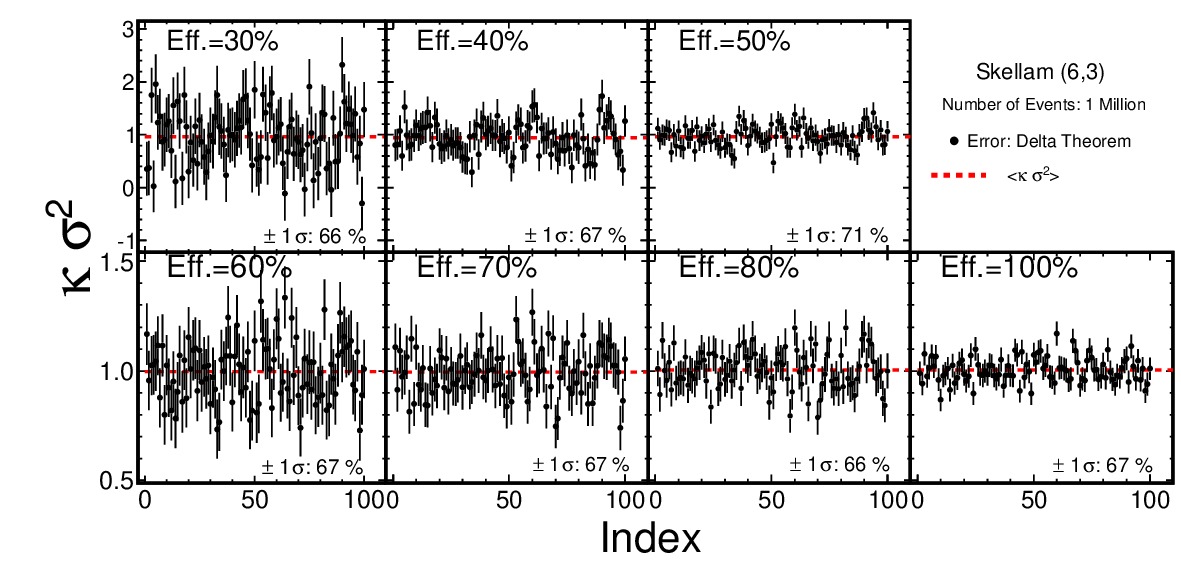}
\caption[]{(Color online) Each data point in each panel represents the efficiency corrected {\KV} and statistical error for an event sample with one million events that independently and randomly generated from the original skellam distribution with efficiency effects. Different panels are with different efficiency varying from 30\% to 100\% The error estimation is based on the Delta theorem. The dashed line in each panel is the average {\KV} value of the 100 samples.   } \label{fig:KV_eff}
\end{center}\end{figure}
\begin{figure}[htbp]
\begin{center}
\hspace{-1.5cm}
\includegraphics[scale=0.75]{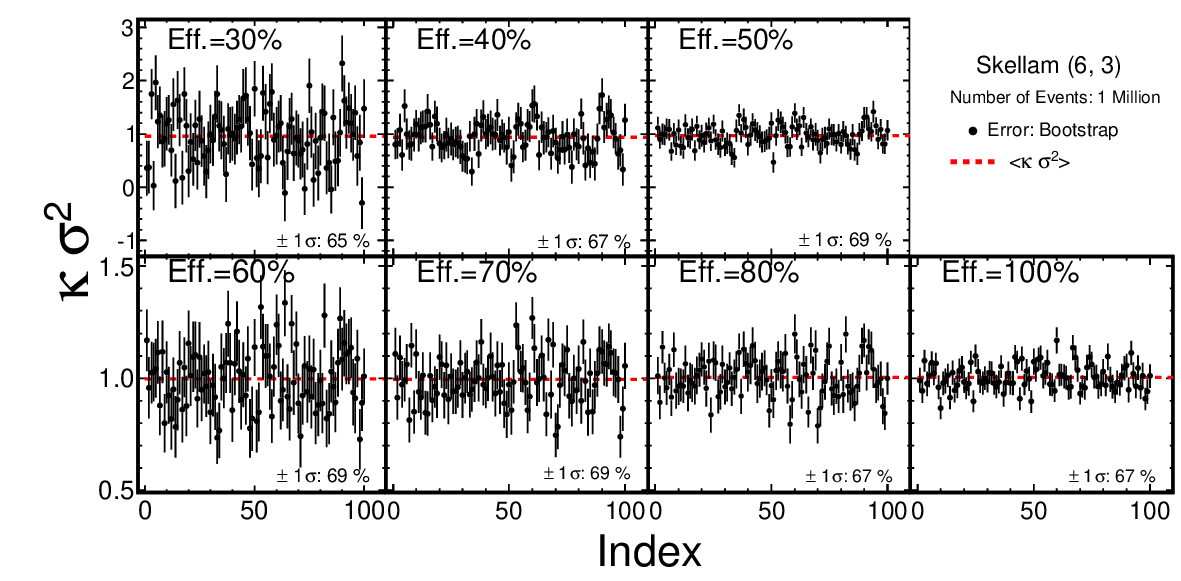}
\caption[]{(Color online) Each data point in each panel represents the efficiency corrected {\KV} and statistical error for an event sample with one million events that independently and randomly generated from the original skellam distribution with efficiency effects. Different panels are with different efficiency varying from 30\% to 100\% The error estimation is based on the Bootstrap. The dashed line in each panel is the average {\KV} value of the 100 samples.     } \label{fig:KV_eff_Boot}
\end{center}\end{figure}

Figure \ref{fig:KV_eff} shows the MC simulation results of efficiency corrected {\KV} value and corresponding statistical errors, which are calculated by using the formulae~\ref{eq:app:KV} in the appendix. At a given efficiency number in each panel, there are one hundred data points, each of which represents efficiency corrected {\KV} value and statistical error of one event sample. Each event sample consists of one million events generated from the original skellam distribution with efficiency effects as shown in the Fig. \ref{fig:dis}. The efficiency numbers are varied from 30\% to 100\% (see each panel in Fig. \ref{fig:KV_eff}). Based on the statistic theory,  if the statistical errors of {\KV} are correctly calculated and can reflect the gaussian statistical fluctuations, the probability for the error bar of data points touching the mean value in each panel should be 68\%. In the Fig. \ref{fig:KV_eff}, this probability for the studied seven cases are 66\%-71\% and the average value for the seven cases is about 67.3\%, which is very close to the theoretical expectation 68\%. The MC simulation results indicate the statistical errors estimated from Delta theorem for efficiency corrected {\KV} based on original skellam distributions are correct and can reasonably reflect the statistical fluctuations for different efficiency values. 

\begin{figure}[htbp]
\begin{center}
\hspace{-1.5cm}
\includegraphics[scale=0.75]{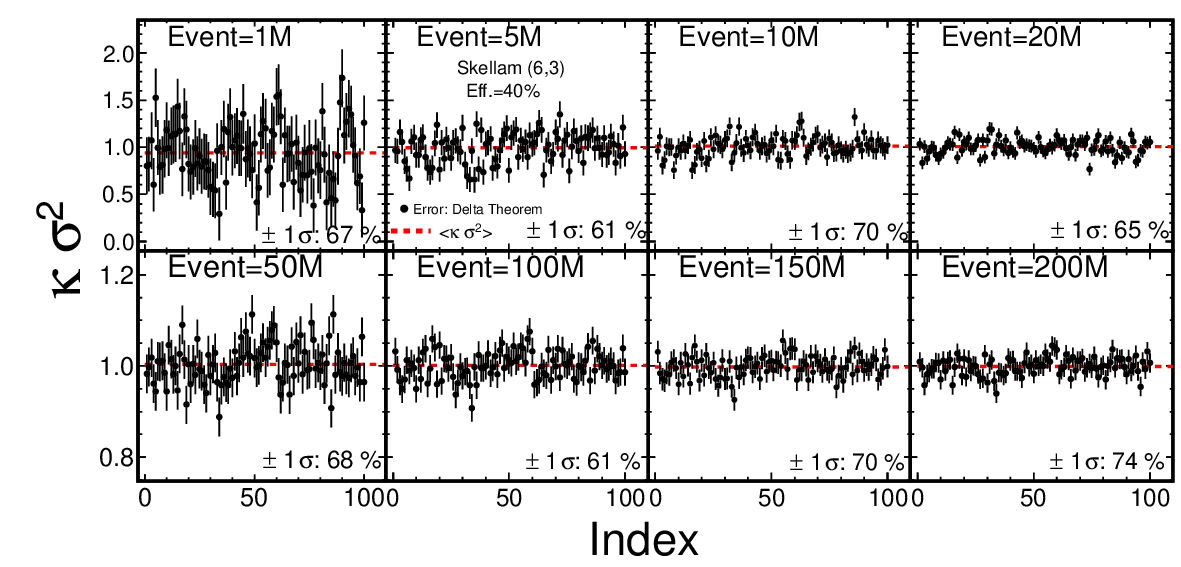}
\caption[]{(Color online) Each data point in each panel represents the efficiency corrected {\KV} and statistical error for an event sample that independently and randomly generated from the original skellam distribution with efficiency effects ($\varepsilon=0.4$). Samples in different panels are with a different number of events varying from 1 to 200 million.  The error estimation is based on the Delta theorem. The dashed line in each panel is the average {\KV} value of the 100 samples.    } \label{fig:KV_events}
\end{center}\end{figure}

Besides the Delta theorem for estimating the statistical errors described in the section 3, another computer intensive one is so called Bootstrap, which is based on resampling methods. For Bootstrap method, one makes $B$ new samples from the original sample of $N$ events. Each of these samples are also with $N$ events, which are chosen randomly with replacement from the original sample. The uncertainty on a statistic quantity is estimated by the root mean square of the $B$ values of
the statistic quantity obtained from these samples. In the MC simulation, we set the number of new samples $B=200$. The variance 
of the statistic quantity $\Phi$ can be given by
\begin{equation}
\begin{split}
V(\Phi ) &= \frac{{\sum\limits_{b = 1}^B {{{\left( {{\Phi _b} - \frac{1}{B}\sum\limits_{b = 1}^B {{\Phi _b}} } \right)}^2}} }}{{B - 1}}\\
 &= \frac{B}{{B - 1}}\left[ {\frac{1}{B}\sum\limits_{b = 1}^B {\Phi _b^2 - {{\left( {\frac{1}{B}\sum\limits_{b = 1}^B {{\Phi _b}} } \right)}^2}} } \right]
\end{split}
\end{equation}
Since one cannot obtain
events further into the tails than those in the original sample, 
the  bootstrap method might run into difficulties if the quantity whose variance is being
estimated depends heavily on the tails of distributions.  For comparison purpose, we also show the error estimation for the efficiency corrected {\KV} from Bootstrap method in the Fig. \ref{fig:KV_eff_Boot} with the same condition as in the Fig. \ref{fig:KV_eff}. We found that the Bootstrap method can reasonably describe the statistical errors of the efficiency corrected {\KV} with various efficiency numbers. The probability for the error bar of data points touching the mean value is ranging from 65\% to 69\% for all panels and the mean value for the seven cases is about  67.6\%, which is very close to the expected value 68\%.  The data points shown in the Fig. \ref{fig:KV_eff} and \ref{fig:KV_eff_Boot} looks very similar. Actually, as we want to concentrate on the comparison of the statistical error bars calculated from the Delta theorem and Bootstrap methods, the data points are calculated from the same data sets for Fig. \ref{fig:KV_eff} and \ref{fig:KV_eff_Boot} and the values are identical, while the  statistical error bars are evaluated from the Delta theorem (Fig. \ref{fig:KV_eff}) and Bootstrap  (Fig. \ref{fig:KV_eff_Boot}), respectively. The statistical error bars of data points in the two figures are very close, but not identical. This consistency can also verify that the analytical error formulae derived from Delta theorem is correct. However, the calculation speed of Delta theorem method is much faster than that of Bootstrap method.

\begin{figure}[htbp]
\begin{center}
\includegraphics[scale=0.45]{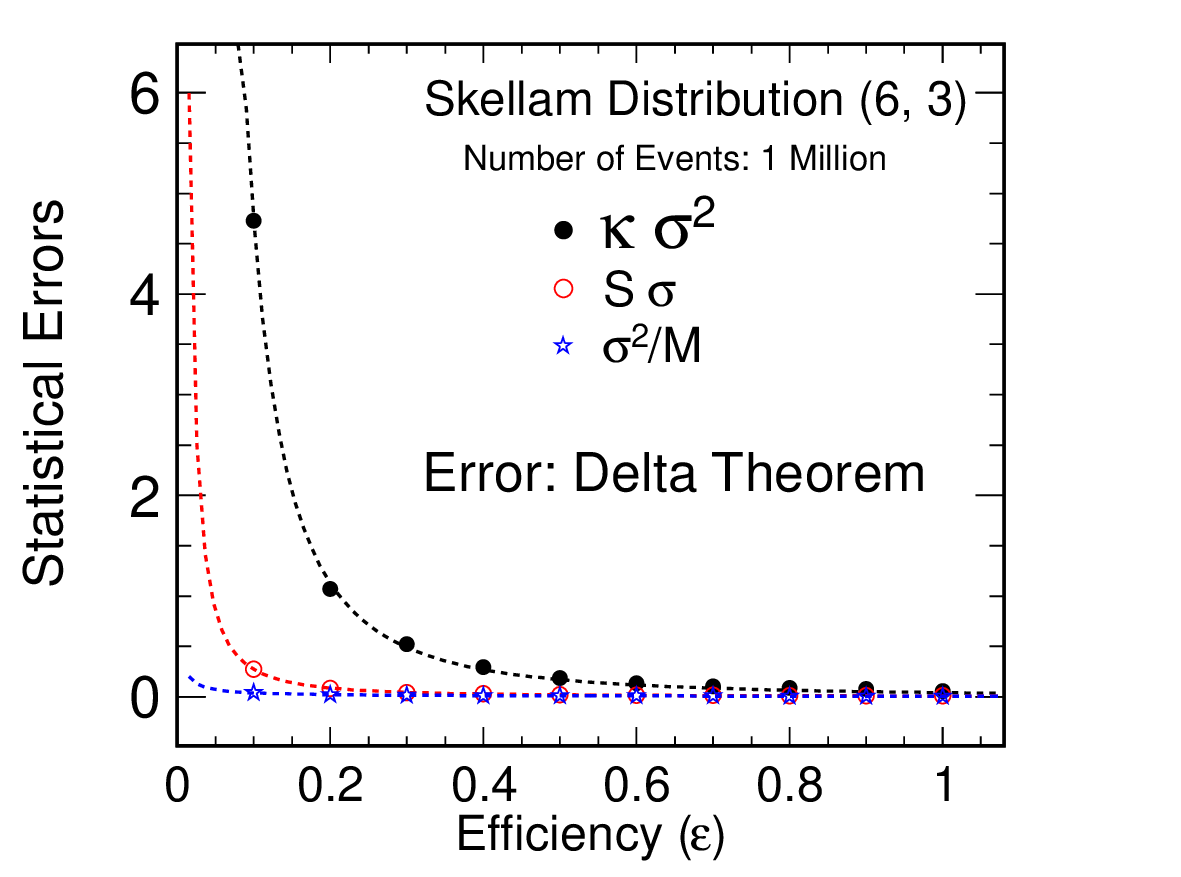}
\caption[]{(Color online) The statistical errors of efficiency corrected {\KV}, {\SD} and {\VM} as a function of efficiency for the original skellam distribution.
The errors are calculated by the Delta theorem.} \label{fig:error_eff}
\end{center}\end{figure}
In Fig. \ref{fig:KV_events}, each data point in each panel represents the efficiency corrected {\KV} value and statistical error of an event sample. Every sample is randomly and independently generated from original skellam distribution with efficiency effects. For samples in the different panels, it is of the same efficiency number 40\% but with different number of events varying from 1 to 200 million. The statistical errors are calculated by the error formulae derived from the error propagation based on the Delta theorem. There are one hundred data points in each panel. It shows that  the probability for the error bar of the data points touching the mean value in the eight panels are from 61\% to 74\% and the average value is 67\%, which is in good agreement with the theoretical expectation 68\%. In addition, when the number of events increases, the {\KV} value and its errors are consistently converged to the expectation value 1.  The simulation results in the Fig. \ref{fig:KV_events} further support the validity of the errors calculated from the Delta theorem for the efficiency corrected moments. Please be note that the scales of Y-axis are different between upper and lower panels for Fig. \ref{fig:KV_eff},  \ref{fig:KV_eff_Boot} and  \ref{fig:KV_events}.

Figure  \ref{fig:error_eff} shows the statistical errors for the efficiency corrected {\KV}, {\SD} and {\VM} as a function of efficiency. In the MC simulation, the efficiency effects are implemented for the original skellam distribution and the number of events is one million for each data point. The statistical errors are dramatically increase when decreasing the efficiency number. This is reasonable since the efficiency number always appears in the denominator of the error formula as shown in the Eq. (\ref{eq:err_var}). These data points can be fitted with functional form:
\begin{equation}
f(\varepsilon ) = \frac{1}{{\sqrt n }}\frac{a}{{{\varepsilon ^b}}}\end{equation}
where $n$ is the number of events ($10^{6}$) which is treated as constant, $a$ and $b$ are free parameters. The fitting results of $a$ and $b$ are 40.6 and 2.06 for {\KV},  6.02 and 1.65 for {\SD}, 4.96 and 0.89 for {\VM}, respectively. The parameters $a$ and $b$ are determined by the original distribution and the studied statistic quantity. One could understand the effects of the efficiency on 
the statistical errors in an intuitive way.  Efficiency will lead to the loss of information of the original distribution, especially tails. Thus,  the smaller the efficiency is, larger uncertainty we will get for the efficiency corrected moments.

\section{Summary}
In this paper, we provide a unified description of efficiency correction and 
error estimation for various order moments of multiplicity distributions in heavy-ion collisions. This description is universal and can be applied to 
the moments analysis for different experiments.  The basic idea is to express the moments and cumulants in terms of the factorial moments, which can be easily corrected for efficiency effects. By knowing the covariance between multivariate factorial moments, we have derived the error formulae for efficiency corrected moments based on the Delta theorem.  To check the validity of the error estimation for different methods, we have done a Monto Carlo simulation based on skellam distribution. The statistical errors obtained from the Delta theorem and Bootstrap methods can reasonably reflect the statistical fluctuations with different efficiency numbers and show consistency converge when increasing the statistics. 

\section*{Acknowledgement}
The work was supported in part by the MoST of China 973-Project No.2015CB856901, NSFC under grant No. 11205067, 11221504.  

\appendix     

\section{Error Estimation for $C_{3}$, $C_{4}$, {\SD} and {\KV} of net-proton distribution.}
The error formulae for the efficiency corrected cumulants and cumulant ratios {\SD} and {\KV} of net-proton distributions will be presented here. For simplify, we provide only the results for the case, where the proton and anti-proton are with constant efficiency $\varepsilon_p$ and  $\varepsilon_{\bar p}$ within the entire phase space, respectively. We define symbols $F_{i,j}$ ( $f_{i,j}$), which are the efficiency corrected (uncorrected) factorial moments for proton and anti-proton distributions. The relation between $F_{i,j}$ and $f_{i,j}$ reads:
\begin{equation}{F_{i,j}} = \frac{{{f_{i,j}}}}{{\varepsilon _p^i\varepsilon _{\bar p}^j}}\end{equation}
The mean value $M$ is defined as: 
\begin{equation}M=F_{1,0}-F_{0,1}\end{equation}
and the covariance between the factorial moments are calculated as:
\begin{equation}
\begin{split}
Cov({f_{r,s}},{f_{u,v}}) &= \frac{1}{n}(\sum\limits_{i = 0}^r {\sum\limits_{j = 0}^s {\sum\limits_{k = 0}^u {\sum\limits_{h = 0}^v {\sum\limits_{\alpha  = 0}^{i + k} {\sum\limits_{\beta  = 0}^{j + h} {[{s_1}(r,i){s_1}(s,j){s_1}(u,k){s_1}(v,h)} } } } } } \\
& \times {s_2}(i + k,\alpha ){s_2}(j + h,\beta ){f_{\alpha ,\beta }}] - {f_{r,s}}{f_{u,v}})
\end{split}
\end{equation}
The differential coefficients $D$ are defined as:
\begin{equation}{D_{i,j}} = \frac{{d\Phi }}{{d{F_{i,j}}}}\end{equation}
where $\Phi$ is statistic quantity.

Firstly, we provide the general formula for the complete derivation of the $r^{th}$ order central moments ($\mu_r$) of net-proton distribution with respect to the joint factorial moments $F_{i,j}$ of proton and anti-proton. 
The central moments $\mu_r$ can be expressed by the factorial moments as:
\begin{equation}
\begin{split}
{\mu _r} &=  < {({N_p} - {N_{\bar p}} - M)^r} > \\
 &= \sum\limits_{s = 0}^r {\left( {\begin{array}{*{20}{c}}
r\\
s
\end{array}} \right)}  < {({N_p} - {N_{\bar p}})^{r - s}} > {( - 1)^s}{M^s}\\
& = \sum\limits_{s = 0}^r {\sum\limits_{t = 0}^{r - s} {\left( {\begin{array}{*{20}{c}}
r\\
s
\end{array}} \right)\left( {\begin{array}{*{20}{c}}
{r - s}\\
t
\end{array}} \right) < N_p^{r - s - t}N_{\bar p}^t > {{( - 1)}^t}{{( - 1)}^s}{M^s}} } \\
& = \sum\limits_{s = 0}^r {\sum\limits_{t = 0}^{r - s} {\sum\limits_{i = 0}^{r - s - t} {\sum\limits_{j = 0}^t {{{( - 1)}^{s + t}}{M^s}\left( {\begin{array}{*{20}{c}}
r\\
s
\end{array}} \right)\left( {\begin{array}{*{20}{c}}
{r - s}\\
t
\end{array}} \right){s_2}(r - s - t,i){s_2}(t,j)} } } } {F_{i,j}}
\end{split}
\end{equation}
Thus, the complete derivation can be written as: 
\begin{equation}  \label{eq:diff_central}
\begin{split}
&\frac{{d{\mu _r}}}{{d{F_{i,j}}}} = \sum\limits_{s = 0}^r {\sum\limits_{t = 0}^{r - s} {{{( - 1)}^{s + t}}{M^s}\left( {\begin{array}{*{20}{c}}
r\\
s
\end{array}} \right)\left( {\begin{array}{*{20}{c}}
{r - s}\\
t
\end{array}} \right){s_2}(r - s - t,i){s_2}(t,j)} } \\
& + \sum\limits_{s = 0}^r {\sum\limits_{t = 0}^{r - s} {\sum\limits_{u = 0}^{r - s - t} {\sum\limits_{v = 0}^t {{{( - 1)}^{s + t + j}}s{M^{s - 1}}{\delta _{1,i + j}}\left( {\begin{array}{*{20}{c}}
r\\
s
\end{array}} \right)\left( {\begin{array}{*{20}{c}}
{r - s}\\
t
\end{array}} \right){s_2}(r - s - t,u){s_2}(t,v)} } } } {F_{u,v}}
\end{split}
\end{equation}
where  $r,s,t,u,v,i,j$ are non-negative integer and $0 < i + j \le r$. If one is dealing with the two sub-phase spaces case, in which the proton and anti-proton are with different efficiency 
in different sub-phase spaces. Then, we need to calculate the derivation of bivariate factorial moments $F_{i,j}$ with respect to the four-variate factorial moments $F_{r,s,u,v}$. 
Based on the Eq. (\ref{eq:4Fto2F}), we have:
\begin{equation}
\frac{{d{F_{i,j}}}}{{d{F_{r,s,u,v}}}} = \sum\limits_{a = 0}^i {\sum\limits_{b = 0}^j {\sum\limits_{c = 0}^a {\sum\limits_{d = 0}^b {{s_1}(i,a)} } } } {s_1}(j,b)\left( {\begin{array}{*{20}{c}}
a\\
c
\end{array}} \right)\left( {\begin{array}{*{20}{c}}
b\\
d
\end{array}} \right){s_2}(a - c,r){s_2}(c,s){s_2}(b - d,u){s_2}(d,v)
\end{equation}
The complete differential for {\KV} of net-proton distribution with respect to the factorial moments:
\begin{equation}  \label{eq:dkvdF}
{D_{i,j}} = \frac{{d(\kappa {\sigma ^2})}}{{d{F_{i,j}}}} = \frac{{d({\mu _4}/{\mu _2} - 3{\mu _2})}}{{d{F_{i,j}}}} = \frac{1}{{{\mu _2}}}\frac{{d{\mu _4}}}{{d{F_{i,j}}}} - (\frac{{{\mu _4}}}{{\mu _2^2}} + 3)\frac{{d{\mu _2}}}{{d{F_{i,j}}}}
\end{equation}
where the two complete differentials $d{\mu _4}/d{F_{i,j}}$ and $d{\mu _2}/d{F_{i,j}}$ can be calculated by using Eq. (\ref{eq:diff_central}). The $r.h.s.$ of Eq.(\ref{eq:dkvdF}) can be expressed in terms of the factorial moments.  For the two sub-phase spaces case, we have:
\begin{equation}{D_{r,s,u,v}} = \sum\limits_{i = 0,j = 0}^4 {\frac{{d({\mu _4}/{\mu _2} - 3{\mu _2})}}{{d{F_{i,j}}}}\frac{{d{F_{i,j}}}}{{d{F_{r,s,u,v}}}}}  = \sum\limits_{i = 0,j = 0}^4 {\left( {\frac{1}{{{\mu _2}}}\frac{{d{\mu _4}}}{{d{F_{i,j}}}} - (\frac{{{\mu _4}}}{{\mu _2^2}} + 3)\frac{{d{\mu _2}}}{{d{F_{i,j}}}}} \right)\frac{{d{F_{i,j}}}}{{d{F_{r,s,u,v}}}}}  \end{equation}
where the condition $i+j\le4$ should be satisfied.

\subsection{Third Order Cumulant ($C_{3}$)}
The $3^{rd}$ order cumulant of net-proton distribution can be expressed in terms of the joint factorial moments of proton and anti-proton distributions.
\begin{equation}
\begin{split}
{C_3}({N_p} - {N_{\bar p}})& = 2{M^3} + M\left( { - 3{F_{0,1}} - 3{F_{0,2}} - 3{F_{1,0}} + 6{F_{1,1}} - 3{F_{2,0}} + 1} \right)\\
 &- 3{F_{0,2}} - {F_{0,3}} + 3{F_{1,2}} + 3{F_{2,0}} - 3{F_{2,1}} + {F_{3,0}}
\end{split}
\end{equation}
The differential coefficients $D$ can be calculated as:
\begin{equation}{D_{i,j}} = \frac{{d{\mu _3}}}{{d{F_{i,j}}}}\end{equation}
One has:
\begin{equation}
\begin{split}
&{D_{3,0}} = 1,{D_{2,1}} =  - 3,{D_{1,2}} = 3,{D_{0,3}} =  - 1\\
&{D_{2,0}} = 3 - 3M,{D_{1,1}} = 6M,{D_{0,2}} = 3 - 3M\\
&{D_{1,0}} =  - 3{F_{0,1}} - 3{F_{0,2}} - 3{F_{1,0}} + 6{F_{1,1}} - 3{F_{2,0}} + 6{M^2} - 3M + 1\\
&{D_{0,1}} = 3{F_{0,1}} + 3{F_{0,2}} + 3{F_{1,0}} - 6{F_{1,1}} + 3{F_{2,0}} - 6{M^2} - 3M - 1
\end{split}
\end{equation}
For other subscripts, the differential coefficients $D$ are zero and the variance of the third order cumulants is calculated as:
\begin{equation}
V({C_3}({N_p} - {N_{\bar p}})) = \sum\limits_{r,s = 0}^3 {\sum\limits_{u,v = 0}^3 {\frac{{{D_{r,s}}}}{{\varepsilon _p^{r + s}}}\frac{{{D_{u,v}}}}{{\varepsilon _{\bar p}^{u + v}}}} } Cov({f_{r,s}},{f_{u,v}})
\end{equation}
When $\varepsilon_p=\varepsilon_{\bar p}=1$,  then $F_{i,j}=f_{i,j}$, we have: 
\begin{equation}
V({C_3}({N_p} - {N_{\bar p}}))= \frac{1}{n}({\mu _6} - 6{\mu _4}{\mu _2} - \mu _3^2 + 9\mu _2^3)
\end{equation}
where $u_{r}$ is the $r^{th}$ order central moments of net-proton distributions.
\subsection{Forth Order Cumulant ($C_{4}$)}
The $4^{th}$ order cumulant of net-proton distribution can be expressed in terms of joint factorial moments of proton and anti-proton distributions.
\begin{equation}
\begin{split}
{C_4}({N_p} - {N_{\bar p}})& = 6{M^2}\left( {{F_{0,1}} + {F_{0,2}} + {F_{1,0}} - 2{F_{1,1}} + {F_{2,0}}} \right)\\
 &- 3{\left( {{F_{0,1}} + {F_{0,2}} + {F_{1,0}} - 2{F_{1,1}} + {F_{2,0}} - {M^2}} \right)^2}\\
 &- 4M\left( { - 3{F_{0,2}} - {F_{0,3}} + 3{F_{1,2}} + 3{F_{2,0}} - 3{F_{2,1}} + {F_{3,0}} + M} \right)\\
 &+ {F_{0,1}} + 7{F_{0,2}} + 6{F_{0,3}} + {F_{0,4}} + {F_{1,0}} - 4\left( {{F_{1,1}} + 3{F_{1,2}} + {F_{1,3}}} \right)\\
 &+ 7{F_{2,0}} + 6\left( {{F_{1,1}} + {F_{1,2}} + {F_{2,1}} + {F_{2,2}}} \right) + 6{F_{3,0}}\\
 &- 4\left( {{F_{1,1}} + 3{F_{2,1}} + {F_{3,1}}} \right) + {F_{4,0}} - 3{M^4}
\end{split}
\end{equation}
The differential coefficients $D$ can be calculated as:
\begin{equation}{D_{i,j}} = \frac{{d({\mu _4} - 3\mu _2^2)}}{{d{F_{i,j}}}} = \frac{{d{\mu _4}}}{{d{F_{i,j}}}} - 6{\mu _2}\frac{{d{\mu _2}}}{{d{F_{i,j}}}} \end{equation}

Thus, we have:
\begin{equation}
\begin{split}
{D_{4,0}} &= 1,{D_{3,1}} =  - 4,{D_{2,2}} = 6,{D_{1,3}} =  - 4,{D_{0,4}} = 1\\
{D_{3,0}} &= 6 - 4M,{D_{2,1}} =  - 6 + 12M,{D_{1,2}} =  - 6 - 12M,{D_{0,3}} = 6 + 4M\\
{D_{2,0}} &=  - 6\left( {{F_{0,1}} + {F_{0,2}} + {F_{1,0}} - 2{F_{1,1}} + {F_{2,0}} - {M^2}} \right) + 6{M^2} - 12M + 7\\
{D_{1,1}} &= 12\left( {{F_{0,1}} + {F_{0,2}} + {F_{1,0}} - 2{F_{1,1}} + {F_{2,0}} - {M^2}} \right) - 12{M^2} - 2\\
{D_{0,2}} &=  - 6\left( {{F_{0,1}} + {F_{0,2}} + {F_{1,0}} - 2{F_{1,1}} + {F_{2,0}} - {M^2}} \right) + 6{M^2} + 12M + 7\\
{D_{1,0}} &=  - 6(1 - 2M)\left( {{F_{0,1}} + {F_{0,2}} + {F_{1,0}} - 2{F_{1,1}} + {F_{2,0}} - {M^2}} \right)\\
 &+ 12M\left( {{F_{0,1}} + {F_{0,2}} + {F_{1,0}} - 2{F_{1,1}} + {F_{2,0}}} \right)\\
 &- 4\left( { - 3{F_{0,2}} - {F_{0,3}} + 3\left( {{F_{1,1}} + {F_{1,2}}} \right) + 3{F_{2,0}} - 3\left( {{F_{1,1}} + {F_{2,1}}} \right) + {F_{3,0}} + M} \right)\\
 &- 12{M^3} + 6{M^2} - 4M + 1\\
{D_{0,1}}& =  - 6(2M + 1)\left( {{F_{0,1}} + {F_{0,2}} + {F_{1,0}} - 2{F_{1,1}} + {F_{2,0}} - {M^2}} \right)\\
& - 12M\left( {{F_{0,1}} + {F_{0,2}} + {F_{1,0}} - 2{F_{1,1}} + {F_{2,0}}} \right)\\
& + 4\left( { - 3{F_{0,2}} - {F_{0,3}} + 3\left( {{F_{1,1}} + {F_{1,2}}} \right) + 3{F_{2,0}} - 3\left( {{F_{1,1}} + {F_{2,1}}} \right) + {F_{3,0}} + M} \right)\\
& + 12{M^3} + 6{M^2} + 4M + 1
\end{split}
\end{equation}
The variance of the forth order cumulants can be written as:
\begin{equation}V({C_4}({N_p} - {N_{\bar p}})) = \sum\limits_{r,s = 0}^4 {\sum\limits_{u,v = 0}^4 {\frac{{{D_{r,s}}}}{{\varepsilon _p^{r + s}}}\frac{{{D_{u,v}}}}{{\varepsilon _{\bar p}^{u + v}}}} } Cov({f_{r,s}},{f_{u,v}})\end{equation}
When $\varepsilon_p=\varepsilon_{\bar p}=1$,   we obtain: 
\begin{equation}
\begin{split}
V({C_4}({N_p} - {N_{\bar p}})) = \frac{1}{n}({\mu _8} - 12{\mu _6}{\mu _2} - 8{\mu _5}{\mu _3} + 48{\mu _4}\mu _2^2 - \mu _4^2 + 64\mu _3^2{\mu _2} - 36\mu _2^4)
\end{split}
\end{equation}
where $u_{r}$ is the $r^{th}$ order central moments of net-proton distributions.

\subsection{Cumulant Ratio $C_3/C_2$ ({\SD})}
We express the {\SD} of net-proton distribution in terms of joint factorial moments of proton and anti-proton distributions.
\begin{equation}
\begin{split}
S\sigma  = \frac{{{C_3}}}{{{C_2}}} = \frac{{{\mu _3}}}{{{\mu _2}}} 
\end{split}
\end{equation}
The differential coefficients $D$ can be calculated as :
\begin{equation}
{D_{i,j}} = \frac{{d({\mu _3}/{\mu _2})}}{{d{F_{i,j}}}} = \frac{1}{{{\mu _2}}}\frac{{d {\mu _3}}}{{d {F_{i,j}}}} - \frac{{{\mu _3}}}{{\mu _2^2}}\frac{{d{\mu _2}}}{{d{F_{i,j}}}}
\end{equation}
Then, we can obtain:
\begin{equation}
\begin{split}
{D_{3,0}} &= 1/{\mu _2},{D_{2,1}} =  - 3/{\mu _2},{D_{1,2}} = 3/{\mu _2},{D_{0,3}} =  - 1/{\mu _2}\\
{D_{2,0}} &= (3 - 3M)/{\mu _2},{D_{1,1}} = 6M/{\mu _2} + 2{\mu _3}/\mu _2^2,{D_{0,2}} = ( - 3 - 3M)/{\mu _2} - {\mu _3}/\mu _2^2\\
{D_{1,0}} &= [ - 3\left( {{F_{0,1}} + {F_{0,2}} + {F_{1,0}} - 2{F_{1,1}} + {F_{2,0}}} \right) + 6{M^2} - 3M + 1]/{\mu _2} - (1 - 2M){\mu _3}/\mu _2^2\\
{D_{0,1}} &= [3\left( {{F_{0,1}} + {F_{0,2}} + {F_{1,0}} - 2{F_{1,1}} + {F_{2,0}}} \right) - 6{M^2} - 3M - 1]/{\mu _2} - (1 + 2M){\mu _3}/\mu _2^2
\end{split}
\end{equation}
The variance of the {\SD} can be calculated as:
\begin{equation}
\begin{split}
V(S\sigma ({N_p} - {N_{\bar p}})) = \sum\limits_{r,s = 0}^3 {\sum\limits_{u,v = 0}^3 {\frac{{{D_{r,s}}}}{{\varepsilon _p^{r + s}}}\frac{{{D_{u,v}}}}{{\varepsilon _{\bar p}^{u + v}}}} } Cov({f_{r,s}},{f_{u,v}})
\end{split}
\end{equation}
When $\varepsilon_p=\varepsilon_{\bar p}=1$,  then $F_{i,j}=f_{i,j}$, we have: 
\begin{equation}
V(S\sigma ({N_p} - {N_{\bar p}})) = \frac{1}{n}(\frac{{{\mu _6}}}{{\mu _2^2}} - \frac{{2{\mu _5}{\mu _3}}}{{\mu _2^3}} + \frac{{{\mu _4}\mu _3^2}}{{\mu _2^4}} + 6\frac{{\mu _3^2}}{{\mu _2^2}} - 6\frac{{{\mu _4}}}{{\mu _2^{}}} + 9{\mu _2})
\end{equation}
where $u_{r}$ is the $r^{th}$ order central moments of net-proton distributions.
\subsection{Cumulant Ratio $C_4/C_2$({\KV})}
We express the {\KV} of net-proton distribution in terms of joint factorial moments of proton and anti-proton distributions.
\begin{equation}
\begin{split}
\kappa {\sigma ^2}& = \frac{{{C_4}}}{{{C_2}}} = \frac{{{\mu _4} - 3\mu _2^2}}{{{\mu _2}}} 
\end{split}
\end{equation}
The differential coefficients $D$ can be calculated as :
\begin{equation}
{D_{i,j}} = \frac{{d({\mu _4}/{\mu _2} - 3{\mu _2})}}{{d {F_{i,j}}}} = \frac{1}{{{\mu _2}}}\frac{{d {\mu _4}}}{{d{F_{i,j}}}} - (\frac{{{\mu _4}}}{{\mu _2^2}} + 3)\frac{{d {\mu _2}}}{{d{F_{i,j}}}}
\end{equation}
Then, one has:
\begin{equation}
\begin{split}
{D_{4,0}}& = 1/{\mu _2},{D_{3,1}} =  - 4/{\mu _2},{D_{2,2}} = 6/{\mu _2},{D_{1,3}} =  - 4/{\mu _2},{D_{0,4}} = 1/{\mu _2}\\
{D_{3,0}} &= (6 - 4M)/{\mu _2},{D_{2,1}} = ( - 6 + 12M)/{\mu _2},{D_{1,2}} = ( - 6 - 12M)/{\mu _2},\\
{D_{0,3}} &= (6 + 4M)/{\mu _2},{D_{2,0}} = (7 - 12M + 6{M^2})/{\mu _2} - ({\mu _4}/\mu _2^2 + 3),\\
{D_{1,1}} &= ( - 2 - 12{M^2})/{\mu _2} + 2({\mu _4}/\mu _2^2 + 3)\\
{D_{0,2}} &= (7 + 12M + 6{M^2})/{\mu _2} - ({\mu _4}/\mu _2^2 + 3)\\
{D_{1,0}} &= (12M{F_{0,1}} + 12M{F_{0,2}} + 12M{F_{1,0}} - 24M{F_{1,1}} + 12M{F_{2,0}} + 12{F_{0,2}}\\
 &+ 4{F_{0,3}} - 12{F_{1,2}} - 12{F_{2,0}} + 12{F_{2,1}} - 4{F_{3,0}} - 12{M^3} + 6{M^2} - 8M + 1)/{\mu _2}\\
& - (1 - 2M)({\mu _4}/\mu _2^2 + 3)\\
{D_{0,1}} &= ( - 12M{F_{0,1}} - 12M{F_{0,2}} - 12M{F_{1,0}} + 24M{F_{1,1}} - 12M{F_{2,0}} - 12{F_{0,2}}\\
& - 4{F_{0,3}} + 12{F_{1,2}} + 12{F_{2,0}} - 12{F_{2,1}} + 4{F_{3,0}} + 12{M^3} + 6{M^2} + 8M + 1)/{\mu _2}\\
& - (1 + 2M)({\mu _4}/\mu _2^2 + 3)
\end{split}
\end{equation}
The variance of the {\KV} can be calculated as:
\begin{equation}  \label{eq:app:KV}
V(\kappa {\sigma ^2}({N_p} - {N_{\bar p}})) = \sum\limits_{r,s = 0}^4 {\sum\limits_{u,v = 0}^4 {\frac{{{D_{r,s}}}}{{\varepsilon _p^{r + s}}}\frac{{{D_{u,v}}}}{{\varepsilon _{\bar p}^{u + v}}}} } Cov({f_{r,s}},{f_{u,v}}) \end{equation}
When $\varepsilon_p=\varepsilon_{\bar p}=1$,  then $F_{i,j}=f_{i,j}$, we have: 
\begin{equation}
\begin{split}
V(\kappa {\sigma ^2}({N_p} - {N_{\bar p}})) &= \frac{1}{n}(\frac{{{\mu _8}}}{{\mu _2^2}} - 6\frac{{{\mu _6}}}{{\mu _2^{}}} - 2\frac{{{\mu _4}{\mu _6}}}{{\mu _2^3}} + 9\mu _4^{} - 8\frac{{{\mu _3}{\mu _5}}}{{\mu _2^2}} + 8\frac{{\mu _3^2{\mu _4}}}{{\mu _2^3}}\\
& + 40\frac{{\mu _3^2}}{{\mu _2^{}}} + \frac{{\mu _4^3}}{{\mu _2^4}} + 6\frac{{\mu _4^2}}{{\mu _2^2}} - 9\mu _2^2)
\end{split}
\end{equation}
where $u_{r}$ is the $r^{th}$ order central moments of net-proton distributions and $n$ is the number of events.

\section*{Reference}
\bibliography{MomentError}
\bibliographystyle{unsrt}
\newpage

\end{document}